\documentclass[12pt]{elsarticle}                                          
\usepackage{axodraw4j}
\usepackage{hyperref}
\usepackage{graphicx}                                                         
\usepackage{a41}                                                                
\usepackage{xcolor}                     
\usepackage{slashed}
\usepackage[rflt]{floatflt}
\usepackage{float}
\usepackage{hyperref}
\hypersetup{colorlinks=true, 
linkcolor=blue, filecolor=blue, urlcolor=mygreen}
\usepackage{breakurl} 
\usepackage{times}    
\usepackage{array}
\usepackage[english]{babel}
\usepackage[latin1]{inputenc}
\usepackage[T1]{fontenc}
\usepackage{ae}
\usepackage{url}
\usepackage{amsmath, amsthm, amssymb}
\usepackage{slashed}

\usepackage{rotating}
\usepackage{graphicx}
\usepackage{comment}
\newcounter{mmacnt}
\def\restartmma{\setcounter{mmacnt}{0}}
\restartmma \catcode`|=\active
\def|#1|{\mathrm{#1}}
\catcode`|=12
\newenvironment{mma}{
\par\smallskip
\catcode`|=\active
\parskip=0pt\parindent=0pt 
\small
\def\In##1\\{%
\def\linebreak{\hfill\break\null\qquad}%
\refstepcounter{mmacnt}
\hangindent=2.5em\hangafter=0
\leavevmode
\llap{\tiny\sffamily In[\arabic{mmacnt}]:=\kern.5em}%
\mathversion{bold}\footnotesiVe$
\displaystyle##1$\normalsiVe
\mathversion{normal}\par
}%
\def\Print##1\\{%
\def\linebreak{\hfill\break}%
\hangindent=2.5em\hangafter=0
\leavevmode ##1\par}%
\def\Out##1\\{%
\def\linebreak{$\hfill\break\null\hfill$}%
\kern\abovedisplayskip\par
\hangindent=2.5em\hangafter=0
\leavevmode
\llap{\tiny\sffamily Out[\arabic{mmacnt}]=\kern.5em}
\footnotesiVe$\displaystyle##1$
\normalsiVe\hfill\null\par
\kern\belowdisplayskip
}%
\def\Warning##1##2\\{%
\def\linebreak{\hfill\break}%
\hangindent=2.5em\hangafter=0
\leavevmode
{\scriptsiVe##1 : ##2}\par}%
}{%
\par\smallskip
}

\usepackage{color}
\newenvironment{fshaded}{%
\MakeFramed {\FrameRestore}
}{\endMakeFramed}

\makeatletter
\def\ps@pprintTitle{%
\let\@oddhead\@empty
\let\@evenhead\@empty
\def\@oddfoot{\reset@font\hfil\thepage\hfil}
\let\@evenfoot\@oddfoot
}
\makeatother
\usepackage{tcolorbox}
\usepackage{slashed}
\begin{document}  
\begin{frontmatter}
\title{\textbf{General Formulas for
Loop-Induced Decays of
$A \to Z\gamma\gamma$ and Their Applications}}
\author[1,2]{Dzung Tri Tran}
\author[3]{Thanh Huy Nguyen}
\author[1,2]{Khiem Hong Phan}
\ead{phanhongkhiem@duytan.edu.vn}
\address[1]{\it Institute of Fundamental
and Applied Sciences, Duy Tan University,
Ho Chi Minh City $70000$, Vietnam}
\address[2]{Faculty of Natural Sciences,
Duy Tan University, Da Nang City $50000$,
Vietnam}
\address[3]{\it
VNUHCM-University of Science,
$227$ Nguyen Van Cu, District $5$,
Ho Chi Minh City, Vietnam}
\pagestyle{myheadings}
\markright{}
\begin{abstract} 
Within the framework of the Standard Model Higgs extensions, including the Two-Higgs-Doublet Model with vector-like fermions and the Triplet-Higgs Model, we derive general one-loop contributions to the rare decay process $A \rightarrow Z \gamma \gamma$. The analytical expressions are formulated with Passarino-Veltman scalar functions, which represent the scalar coefficients of one-loop Lorentz-covariant tensor integrals. These functions are written in accordance with the input conventions of the {\tt LoopTools} and {\tt Collier} packages, facilitating the efficient numerical generation of decay rates and their distributions using these computational tools. As part of our phenomenological study, we examine the branching ratios of this decay channel within the viable parameter space of the considered models. Our results indicate that the branching ratio can reach $\mathcal{O}(10^{-4})$ in the Two-Higgs-Doublet Model and $\mathcal{O}(10^{-2})$ in the Triplet-Higgs Model at specific points within the allowed parameter regions. Furthermore, the inclusion of vector-like fermions in the loop leads to a significant modification of the partial decay rates, with an observed variation of approximately $10\%$. Additionally, we explore the dependence of the branching ratios on key model parameters, including the charged Higgs mass, mixing angles, and the soft-breaking scale, providing deeper insights into the phenomenological implications of these Higgs-extended frameworks.
\end{abstract}
\end{frontmatter}
\section{Introduction}
The scalar Higgs sector of the Standard Model (SM) of particle physics remains one of the least understood parts in both theoretical and experimental research. Consequently, the exploration of the scalar Higgs sector has become a primary objective at future high-energy frontier colliders, such as the High-Luminosity Large Hadron Collider (HL-LHC) and future lepton colliders (LC). From the theoretical side, it is well-established that the scalar potential is extended by including scalar singlets or scalar multiplets in theories beyond the Standard Model (BSM). Many additional scalar particles, such as CP-even Higgs, CP-odd Higgs, and charged Higgs, are predicted in BSMs. From the experimental perspective, searches for additional scalar particles in many BSMs are therefore scrutinized in the exploration of the scalar potential at future colliders. Recent studies have reported the direct searches for
CP-odd Higgs ($A$) boson in the Tevatron and the LHC, as detailed in Ref.~\cite{Dermisek:2009fd}. Searches for the decay $A\rightarrow Zh$
have been conducted at the LHC in Ref.~\cite{ATLAS:2015kpj}.
Additionally, hunts for $A\rightarrow
\mu\mu\tau\tau$ at the LHC in
Refs.~\cite{CMS:2020ffa,ATLAS:2024rzd}
have been reported, while investigations into
$A\rightarrow \mu^+\mu^-$ in $pp$ collisions at $\sqrt{s}=7$
TeV have been discussed in Ref.~\cite{CMS:2012fgd}.
More recently, the decay channel
$A\rightarrow ZH$ with final states $ {\ell}^{+}
{\ell}^{-}t\overline{t} $ and 
$\nu \overline{\nu}b\overline{b}$ at ATLAS detector 
has been measured in Ref.~\cite{ATLAS:2023zkt}.

Detailed theoretical evaluations of one-loop and higher-loop corrections to the decay rates and production cross-sections are essential for
testing consistency with high-precision measurements at future colliders. One-loop contributions to the two-body decays of CP-odd Higgs bosons within the framework of the Two Higgs Doublet Model (THDM), including its CP-conserving versions, have been computed in Refs.~\cite{Aiko:2022gmz, Kanemura:2017gbi, Kanemura:2019slf,
Aiko:2023xui}. One-loop corrections to the important decay channel $A \rightarrow Zh$ have been investigated in Refs.~\cite{Akeroyd:2024tbp, Akeroyd:2023kek}. Likewise, the one-loop corrected decay of $A$ into a pair of scalar fermions has been computed in Refs.~\cite{Weber:2003tw, Weber:2003eg} within the framework of the minimal supersymmetric extension of the Standard Model (MSSM). By employing the Pad\'e improvement method, enhanced predictions for the decay rate of the CP-odd Higgs boson into two gluons were presented in Ref.~\cite{Chishtie:1999dd}. Phenomenological investigations of the decay of a CP-odd scalar into gauge bosons in the THDM with a vector-like quark (VLQ) at the LHC were explored in Ref.~\cite{Arhrib:2018pdi}. The distinction between the pseudoscalar decay mode $A \rightarrow ZH$ and the CP-even scalar decay $H \rightarrow ZA$ in $Zt\bar{t}$ final states at the LHC, based on top-quark spin correlations, was examined in Ref.~\cite{Arco:2025ydq}. The correlation between the CP-odd Higgs decays into diphotons and electric dipole moments within the THDM was systematically analyzed in Ref.~\cite{Banik:2024ugs}. In our previous studies, the decay channels
$A \rightarrow \ell \bar{\ell}V$, where $V = \gamma,Z$, were recently analyzed within the Standard Model Higgs extensions (HESMs), including the THDM and the Triplet Higgs Model (THM); see Ref.~\cite{Phan:2024zus} for further details. Moreover, Ref.~\cite{Tran:2025dea} presented the first analytic expressions for one-loop contributions to the rare decay process $A \rightarrow hh\gamma$ within the CP-conserving THDM. Furthermore, the three-body decay processes of the CP-odd Higgs, involving a $Z$ boson and a diphoton in the final state, have been rigorously computed in Refs.~\cite{Sanchez-Velez:2018xdj, Sanchez-Velez:2019nsh}. The production cross-sections of the CP-odd Higgs boson in association with a neutral $Z$ boson at the LHC have been systematically evaluated within the MSSM in Ref.~\cite{Yin:2002sq}, for the THDM
in Ref.~\cite{Akeroyd:1999gu}, and in the context of supersymmetric models
in Ref.~\cite{Akeroyd:2001aka}. The scattering process $e^-\gamma \rightarrow e^- A$ has been proposed as a potential probe in Ref.~\cite{Sasaki:2017fvk}. The feasibility of investigating CP-odd Higgs production via the process $e^+ e^- \rightarrow \nu \bar{\nu} A$ within the THDM has also been studied in Ref.~\cite{Farris:2003pn}. Additionally, the production processes $e^+ e^- \rightarrow \gamma A,~ZA,~\nu_e \bar{\nu}_e A$ within the MSSM have been computed in Ref.~\cite{Arhrib:2002ti}.

Studying the rare decay $A \rightarrow Z\gamma\gamma$ provides an alternative approach to probing the extended Higgs sector beyond the two-body decays of the CP-odd Higgs boson. If observed, this process would serve as direct evidence of additional Higgs bosons, reinforcing the necessity of an extended scalar sector in various Higgs Extended Scalar Models. Furthermore, unlike tree-level decays, loop-induced processes are highly sensitive to new particles appearing in the loop, such as charged Higgs bosons, vector-like fermions, and other exotic states. The decay rate and kinematic distributions of $A \to Z\gamma\gamma$ can offer valuable constraints
on the exotic mass spectrum and its couplings to gauge bosons. Moreover, the branching ratio of $A \to Z\gamma\gamma$ is strongly dependent on key parameters of HESMs. By analyzing this process, one can impose constraints on the Higgs potential and the electroweak symmetry-breaking mechanism in the considered models.
For these reasons, we present general analytical expressions for one-loop contributions to the rare decay process $A \rightarrow Z \gamma \gamma$, applicable to various the Standard Model Higgs extensions, including the Two-Higgs-Doublet Model with vector-like fermions and the Triplet-Higgs Model. These expressions are formulated using scalar one-loop coefficients, commonly referred to as Passarino-Veltman functions (hereafter denoted as PV-functions), which arise from Lorentz-covariant one-loop tensor integrals~\cite{Denner:1991kt}. The computations adhere to the standard input conventions of the {\tt LoopTools}~\cite{Hahn:1998yk} and {\tt Collier}~\cite{Denner:2016kdg} packages. This approach enables the numerical generation of partial decay rates and their distributions using either of these computational tools. In our phenomenological analysis, we investigate the branching ratios of the computed decay channels within the parameter space of the considered models.

The article is structured as follows: we derive the general one-loop formulas contributing to the decay channel $A \rightarrow Z \gamma \gamma$ in detail and present the explicit analytical expressions for the one-loop form factors in in Section~2.
Section~3 discusses the phenomenological results of our calculations. Finally, Section~4 provides conclusions and future prospects. The appendices summarize the models considered in this study.
\section{General formulas for one-loop-induced
decay of $A \rightarrow Z\gamma \gamma$}
We present a thorough analysis of the one-loop
induced decay process $A \rightarrow Z\gamma \gamma$ in this section. Within the general framework of the Standard Model Higgs Extensions
under consideration, additional scalar particles are introduced, including CP-even Higgs bosons ($\phi = h,~H$), where $h$ is identified as the SM-like Higgs in this work; a CP-odd Higgs $A$; and singly and doubly charged Higgs bosons $S_Q$ with charge $Q$
(where $Q=1,~2$ correspond to the singly and doubly charged Higgs bosons). Beyond these additional scalar particles, many HESMs also incorporate extra vector-like fermions (VLFs). In our calculation, we denote fermions as $F = f, U_j, D_j$, where $f$ represents the SM fermions, and $U_j, D_j$ denote the VLFs.

The new couplings involving these additional scalars and extra VLFs are parameterized as $g_{\text{vertex}}$, as shown in Tables~\ref{Coupling1} and~\ref{Coupling2}. Specifically, Table~\ref{Coupling1} presents the detailed general couplings relevant to the Two-Higgs-Doublet Model for the process under consideration. The specific modes examined in this study are reviewed comprehensively in Appendix~A. The Yukawa couplings of scalar Higgs bosons with fermions are also provided in Appendix~A. Depending on the type of THDM, the explicit formulas for these Yukawa couplings are summarized in Table~\ref{Z2-assignment}. Additionally, we derive the couplings of vector-like fermions with gauge bosons and scalar particles
in Appendix~B.
\begin{table}[H]
\begin{center}
\begin{tabular}{
c@{\hspace{0.5cm}}
c@{\hspace{1cm}}c}
\hline\hline
Vertices
&
Notations 
& 
Couplings  
\\
\hline
$AZ_{\mu}h$
&
$g_{AZh}
\cdot 
(p_{A}-p_{h})_{\mu}$ 
& $i\left(
\frac{M_Z\; 
c_{\beta-\alpha}}{v}
\right)
\cdot 
(p_{A}-p_{h})_{\mu}$  
\\
\hline
$AZ_{\mu}H$
&
$g_{AZH}
\cdot 
(p_{A}-p_{H})_{\mu}$ 
& 
$-i
\left(
\frac{M_Zs_{\beta-\alpha}}
{v}
\right)
\cdot 
(p_{A}-p_{H})_{\mu}$  
\\
\hline
$AW^{\pm}_{\mu}H^{\mp}$
&
$g_{AW^{\pm}H^{\mp}}\cdot
(p_A-p_{H^\pm})_{\mu}$
&
$i\dfrac{M_W}{v}
\cdot (p_A-p_{H^\pm})_{\mu}$
\\
\hline
$hW_{\mu}^{\pm}W_{\nu}^{\mp}$
&
$g_{hW^{\pm}W^{\mp}}
\cdot g_{\mu\nu}$ 
& 
$i
\left(
\frac{2M_W^2}{v}\;
s_{\beta-\alpha}
\right)
\cdot
g_{\mu\nu}$ 
\\
\hline
$hZ_{\mu}Z_{\nu}$
&
$g_{hZZ}
\cdot g_{\mu\nu}$ 
& 
$i
\left(
\dfrac{2M_Z^2}{v}
s_{\beta-\alpha}
\right)
\cdot
g_{\mu\nu}$ 
\\
\hline
$HZ_{\mu}Z_{\nu}$
&
$g_{HZZ}
\cdot 
g_{\mu\nu}$ 
& 
$i
\left(
\dfrac{2M_Z^2}{v}
c_{\beta-\alpha}
\right)
\cdot
g_{\mu\nu}$  
\\
\hline
$HW_{\mu}^{\pm}W_{\nu}^{\mp}$
&
$g_{HW^{\pm}W^{\mp}}
\cdot 
g_{\mu\nu}$ 
& 
$i
\left(
\frac{2M_W^2}{v}\; 
c_{\beta-\alpha}
\right)
\cdot
g_{\mu\nu}$  
\\
\hline
$hH^{\pm}H^{\mp}$ 
&
$g_{hH^{\pm}H^{\mp}}$ 
& $
-i
\dfrac{
(c_{\alpha-3\beta}+3c_{\alpha+\beta})M_h^2
-4(c_{\alpha+\beta}M^2
+s_{\alpha-\beta}s_{2\beta}M_{H^\pm}^2)
}
{
2vs_{2\beta}
}
$ 
\\
\hline
$HH^{\pm}H^{\mp}$ 
& 
$g_{HH^{\pm}H^{\mp}}$ 
& 
$
-
i\dfrac{
(s_{\alpha-3\beta}
+3s_{\alpha+\beta})M_H^2
-
4(s_{\alpha+\beta}M^2
-c_{\alpha-\beta}s_{2\beta}M_{H^\pm}^2)
}
{
2vs_{2\beta}
}$ 
\\
\hline\hline
\end{tabular}
\end{center}
\caption{
\label{Coupling1}
The general couplings relevant
to the process under consideration
are presented in the context of the THDM.
The detailed derivation of these
couplings can be found
in Refs.~\cite{Hue:2023tdz, Phan:2024vfy, Phan:2024jbx}.
}
\end{table}
The general couplings contributing to
the calculated process in the Triplet-Higgs Model
are detailed in Table~\ref{Coupling2}.
A summary of the model is presented in
Appendix~C, where several relevant
couplings for this decay channel are
also derived
(particularly in the limit $v_{\Delta} \rightarrow 0$).
\begin{table}[H]
\begin{center}
\begin{tabular}{
c@{\hspace{0.5cm}}
c@{\hspace{0.5cm}}c}
\hline\hline
Vertices  &
Notations & 
Couplings \\
\hline
$AZ_{\mu}h$
&
$g_{AZh}
\cdot 
(p_{A}-p_{h})_{\mu}$  
& 
$-i
\dfrac{M_Z(2c_{\beta^0}\; 
s_{\alpha}-c_{\alpha}\; 
s_{\beta^0}) }{
\sqrt{v^2 + 2v_{\Delta}^2} 
}
\cdot 
(p_{A}-p_{h})_{\mu}$ 
\\
\hline
$AZ_{\mu}H$
&
$g_{AZH}\cdot 
(p_{A}-p_{H})_{\mu}$ 
& 
$-i\dfrac{
M_Z
(
2c_{\beta^0}c_{\alpha}
+s_{\alpha}s_{\beta^0}
)
}{ \sqrt{v^2 + 2v_{\Delta}^2} }
\cdot
(p_{A}-p_{H})_{\mu}$ 
\\
\hline
$hW_{\mu}^{\pm}W_{\nu}^{\mp}$
&
$g_{hW^{\pm}W^{\mp}}
\cdot
g_{\mu\nu}$  
& $i\dfrac{2M_W^2}{v}
(c_{\alpha}c_{\beta^{\pm}}
+\sqrt{2}s_{\alpha}s_{\beta^{\pm}})
\cdot
g_{\mu\nu}$
\\
\hline
$HW_{\mu}^{\pm}W_{\nu}^{\mp}$
&
$g_{HW^{\pm}W^{\mp}}
\cdot
g_{\mu\nu}$  
& $i\dfrac{2M_W^2}{v}
(-s_{\alpha}c_{\beta^{\pm}}
+\sqrt{2}c_{\alpha}s_{\beta^{\pm}})
\cdot
g_{\mu\nu}$ 
\\
\hline
$HZ_{\mu}Z_{\nu}$
&
$g_{HZZ}\cdot 
g_{\mu\nu}$
&
$i\dfrac{M_Z^2}
{\sqrt{v^2 + 2 v_{\Delta}^2
}}(2 c_{\alpha}s_{\beta^0}
-s_{\alpha}c_{\beta^0}) 
\cdot 
g_{\mu\nu}$
\\ \hline
$AW^{\pm}_{\mu}H^{\mp}$
&
$g_{AW^{\pm}H^{\mp}}\cdot 
(p_A-p_{H^\pm})_{\mu}$ 
&
$i\dfrac{M_W}{v}
(s_{\beta^\pm}s_{\beta^0}
+\sqrt{2}c_{\beta^\pm}c_{\beta^0})
\cdot (p_A-p_{H^\pm})_{\mu}$ 
\\
\hline
$hH^{\pm}H^{\mp}$
&
$g_{hH^{\pm}H^{\mp}}$ 
& $i\Big[
\dfrac{v^2s_{\alpha}}
{v_{\Delta}(2v_{\Delta}^2+v^2)}M_A^2
-
\dfrac{2(v_{\Phi}c_{\alpha}
+2v_{\Delta}s_{\alpha})}{v^2}M_{H^\pm}^2
$
\\
&&
\\
&&
\hspace{4.5cm}
$
-\dfrac{(2v_{\Delta}^3c_{\alpha}
+v_{\Phi}^3s_{\alpha})}
{v_{\Delta}v_{\Phi}v^2}M_h^2
\Big]$ 
\\
\hline
$HH^{\pm}H^{\mp}$
&
$g_{HH^{\pm}H^{\mp}}$  
& $i\Big[
\dfrac{v^2c_{\alpha}}
{v_{\Delta}(2v_{\Delta}^2+v^2)}M_A^2
+\dfrac{2(v_{\Phi}s_{\alpha}
-2v_{\Delta}c_{\alpha})}{v^2}M_{H^\pm}^2
$
\\
&&
\\
&&
\hspace{4.5cm}
$
+\dfrac{(2v_{\Delta}^3s_{\alpha}
-v_{\Phi}^3c_{\alpha})}
{v_{\Delta}v_{\Phi}v^2}M_H^2
\Big]$
\\
\hline
$hH^{\pm\pm}H^{\mp\mp}$
&
$g_{hH^{\pm\pm}H^{\mp\mp}}$ 
& 
$
i
\Big[
\dfrac{v_{\Phi}(2v_{\Delta}c_{\alpha}
-v_{\Phi}s_{\alpha})}
{v_{\Delta}(2v_{\Delta}^2+v^2)}M_A^2
-\dfrac{s_{\alpha}}{v_{\Delta}}
(M_h^2+2M_{H^{\pm\pm}}^2)
$
\\
&&
\\
&&
\hspace{4.5cm}
$
+\dfrac{4v_{\Phi}(v_{\Phi}s_{\alpha}
-v_{\Delta}c_{\alpha})}
{v_{\Delta}v^2}M_{H^\pm}^2
\Big]$
\\
\hline
$HH^{\pm\pm}H^{\mp\mp}$
&
$g_{HH^{\pm\pm}H^{\mp\mp}}$ 
& 
$i\Big[
-\dfrac{v_{\Phi}(2v_{\Delta}s_{\alpha}
+ v_{\Phi}c_{\alpha})}
{v_{\Delta}(2v_{\Delta}^2+v^2)}M_A^2
-\dfrac{c_{\alpha}}{v_{\Delta}}
(M_H^2+2M_{H^{\pm\pm}}^2)
$
\\
&&
\\
&&
\hspace{4cm}
$
+\dfrac{4v_{\Phi}(v_{\Phi}c_{\alpha}
+v_{\Delta}s_{\alpha})}
{v_{\Delta}v^2}M_{H^\pm}^2
\Big]$\\
\hline 
$h\bar{f}f$
&
$g_{h\bar{f}f}$
& 
$i\dfrac{m_f}{v}
\dfrac{c_{\alpha}}{c_{\beta^\pm}}$
\\
\hline
$H\bar{f}f$
&
$g_{H\bar{f}f}$
& $-i\dfrac{m_f}{v}
\dfrac{s_{\alpha}}{c_{\beta^\pm}}$
\\
\hline
$A\bar{f}f$
&
$g_{A\bar{f}f}$
& 
$\pm 
\left( 
\dfrac{m_f
}{v}
\dfrac{
s_{\beta^0}
}
{
c_{\beta^\pm}
}
\right)
\gamma^5$ 
\; ($-$ for $f \equiv u$;
$+$ for $f \equiv d$, $\ell$)
\\
\hline\hline
\end{tabular}
\end{center}
\caption{\label{Coupling2}
The general couplings relevant
to the process under consideration
are presented in the context of the THM.
Notably, the relation
$v_{\Phi} = \sqrt{v^2 - 2v_{\Delta}^2}$
holds in the presented expressions.
A comprehensive derivation of these
couplings is provided in Ref.~\cite{Hue:2023tdz}.
}
\end{table}

\begin{figure}[htp]
\centering
\includegraphics[width=14cm, height=8cm]
{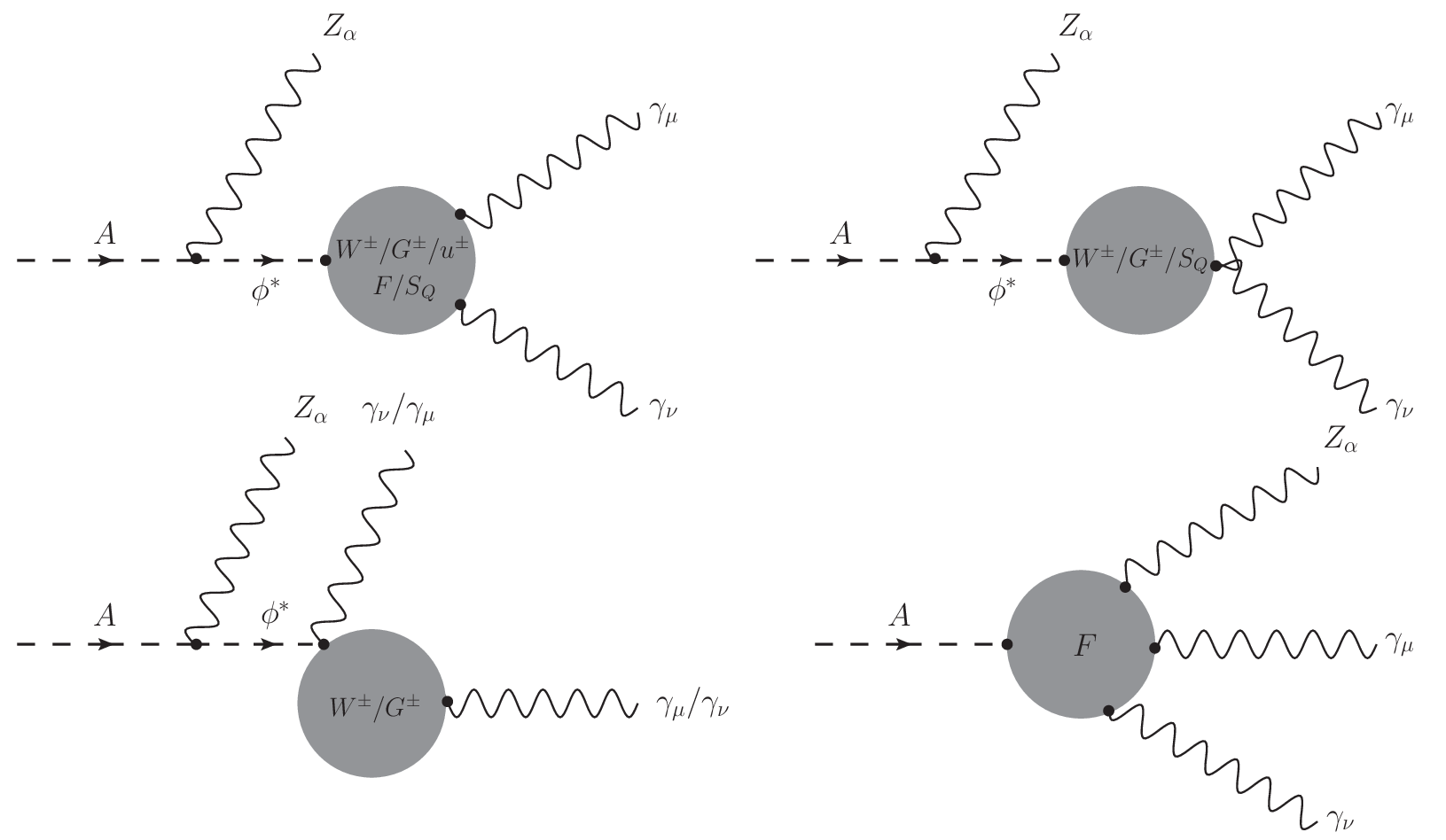}
\caption{
\label{FeyDia-HF}
All one-loop Feynman diagrams contributing to the process
$A \rightarrow Z \gamma \gamma$ are plotted in the HF gauge.
These diagrams include contributions from fermions
(including VLFs), gauge bosons, and charged Higgs bosons. }
\end{figure}

In the HF gauge, all one-loop Feynman diagrams relevant to this process are depicted in Fig.~\ref{FeyDia-HF}. The loop-induced contributions to the processes under investigation can be classified into two distinct topologies. The first topology corresponds to the off-shell decay process $\phi \rightarrow \gamma \gamma$, connected by the $AZ\phi$ coupling. In this category, the loop contributions involve fermions (including VLFs), $W$ gauge bosons along with their associated Goldstone bosons and ghost particles in the context of the HF gauge, as well as charged Higgs bosons $S_Q$. The second topology consists of one-loop diagrams with four external legs, where only fermions and VLFs contribute within the loop.

The HESMs are initially implemented into the computational package {\tt FeynArts}~\cite{Hahn:2000kx}. The calculations are conducted in the 't Hooft-Feynman gauge (HF) with the assistance of {\tt FormCalc}~\cite{Hahn:2016ebn}. In the subsequent discussion, we provide a detailed derivation of the calculations of the process $A \rightarrow Z \gamma \gamma$.
The one-loop induced amplitude for the proces
$A (p) \rightarrow Z_\alpha (k_1) \, \gamma_\mu (k_2) \, \gamma_\nu (k_3)$ can, in general, be expressed through
the following Lorentz structures:
\begin{eqnarray}
\mathcal{M}_{A\rightarrow Z\gamma \gamma}
&=&
\mathcal{M}^{\alpha \mu \nu} \,
\varepsilon^*_\alpha (k_1) \,
\varepsilon^*_\mu (k_2) \,
\varepsilon^*_\nu (k_3).
\end{eqnarray}
Here, $\varepsilon^*_\mu (k_j)$ for $j=1,2,3$
represents for the polarization vectors
corresponding to the neutral gauge bosons
in the final state. The one-loop amplitude is
expressed in kinematic invariant masses,
defined as $p^2 = M_{A}^2,~k_1^2 = M_Z^2,~k_2^2 = k_3^2 = 0$
and $k_{ij} = (k_i + k_j)^2,~\text{for}~i<j,~\text{with}~i,j=1,2,3.$
The Lorentz structure of
$\mathcal{M}_{\mu\nu\alpha}$, parameterized
in terms of $36$ basic form factors,
is given by:
\begin{eqnarray}
\mathcal{M}^{\alpha \mu \nu}
&=&
\sum \limits_{i = 1}^3
\Big[
\mathcal{F}_{1i} \,
k_i^\alpha 
g^{\mu \nu}
+
\mathcal{F}_{2i} \,
k_i^\mu
g^{\alpha \nu}
+
\mathcal{F}_{3i} \,
k_i^\nu 
g^{\mu \alpha}
\Big]
+
\sum \limits_{i, j, l = 1}^3
\mathcal{F}_{i j l} \,
k_{i}^\alpha
k_{j}^\mu
k_{l}^\nu.
\end{eqnarray}
Due to the transversality condition
imposed on the external vector bosons,
the following relation holds:
\begin{eqnarray}
k_1^\alpha \,
\varepsilon^*_\alpha (k_1)
=
k_2^\mu \,
\varepsilon^*_\mu (k_2)
=
k_3^\nu \,
\varepsilon^*_\nu (k_3)
=0. 
\end{eqnarray}
All terms proportional to $k_1^\alpha$, $k_2^\mu$,
and $k_3^\nu$ in the invariant amplitude can be disregarded.
Consequently, the amplitude is simplified and can
be expressed using $14$ form factors, given by:
\begin{eqnarray}
\label{sym1AMP}
\mathcal{M}^{\alpha \mu \nu}
&=&
(
\mathcal{F}_{12} \,
k_2^\alpha
+
\mathcal{F}_{13} \,
k_3^\alpha  
) g^{\mu \nu}
+
(
\mathcal{F}_{21} \,
k_1^\mu
+
\mathcal{F}_{23} \,
k_3^\mu
)
g^{\alpha \nu}
+
(
\mathcal{F}_{31} \,
k_1^\nu 
+
\mathcal{F}_{32} \,
k_2^\nu 
)
g^{\mu \alpha}
\nonumber  \\
&&
+
(
\mathcal{F}_{211} \,
k_2^\alpha
+
\mathcal{F}_{311} \,
k_3^\alpha
) k_1^\mu k_1^\nu
+
(
\mathcal{F}_{212} \,
k_2^\alpha
+
\mathcal{F}_{312} \,
k_3^\alpha
) k_1^\mu k_2^\nu
\nonumber\\
&&
+
(
\mathcal{F}_{231} \,
k_2^\alpha
+
\mathcal{F}_{331} \,
k_3^\alpha
)
k_3^\mu
k_1^\nu
+
( 
\mathcal{F}_{232} \, k_2^\alpha
+
\mathcal{F}_{332} \, k_3^\alpha
)
k_3^\mu
k_2^\nu.
\end{eqnarray}
Furthermore, since the final-state photons are on-shell,
the amplitude satisfies the Ward identity.
Specifically, the amplitude must vanish when
substituting either $\varepsilon^*_\mu (k_2)
\rightarrow k_2^\mu$ or
$\varepsilon^*_\nu (k_3) \rightarrow k_3^\nu$.
Consequently, the following relations among
the form factors are derived:
\begin{eqnarray}
\mathcal{F}_{12}
&=&
-
( k_1 \cdot k_3 )
\,
\mathcal{F}_{231}
-
( k_2 \cdot k_3 )
\,
\mathcal{F}_{232},
\\
\mathcal{F}_{13}
&=&
-
( k_1 \cdot k_2 )
\,
\mathcal{F}_{312}
-
( k_2 \cdot k_3 )
\,
\mathcal{F}_{332},
\\
\mathcal{F}_{21}
&=&
-
( k_1 \cdot k_3 )
\,
\mathcal{F}_{311}
-
( k_2 \cdot k_3 )
\,
\mathcal{F}_{312},
\\
\mathcal{F}_{23}
&=&
-
( k_1 \cdot k_3 )
\,
\mathcal{F}_{331}
+
( k_1 \cdot k_2 )
\,
\mathcal{F}_{312},
\\
\mathcal{F}_{31}
&=&
-
( k_1 \cdot k_2 )
\,
\mathcal{F}_{211}
-
( k_2 \cdot k_3 )
\,
\mathcal{F}_{231},
\\
\mathcal{F}_{32}
&=&
-
( k_1 \cdot k_2 )
\,
\mathcal{F}_{212}
+
( k_1 \cdot k_3 )
\,
\mathcal{F}_{231}.
\end{eqnarray}
By expressing the form factors
$\mathcal{F}_{1i}$, $\mathcal{F}_{2i}$,
and $\mathcal{F}_{3i}$ in terms of
the six-component basis $\mathcal{F}_{ijl}$,
the Lorentz-invariant amplitude
$\mathcal{M}^{\alpha \mu \nu}$ can be
compactly formulated as:
\begin{eqnarray}
\label{invAmp}
\mathcal{M}^{\alpha \mu \nu}
&=&
\dfrac{1}{k_2 \cdot k_3}
\Big[
( k_2 \cdot k_3 )
\,
k_1^{\nu } 
-
( k_1 \cdot k_3 ) 
\,
k_2^{\nu }
\Big] 
\Big[
k_1^{\mu } k_2^{\alpha }
-
( k_1 \cdot k_2 ) 
\,
g^{\alpha \mu }
\Big]
\,
\mathcal{F}_{211}
\nonumber \\
&&\hspace{-0.25cm}
+
\dfrac{1}{k_2 \cdot k_3}
\Big[
( k_2 \cdot k_3 )
\,
k_1^{\mu } 
-
( k_1 \cdot k_2 ) 
\,
k_3^{\mu }
\Big] 
\Big[
k_1^{\nu } k_3^{\alpha }
-
( k_1 \cdot k_3 ) 
\,
g^{\alpha \nu }
\Big]
\,
\mathcal{F}_{311}
\nonumber \\
&&\hspace{-0.25cm}
+ 
\Big\{
\Big[
k_2^{\alpha } k_3^{\mu }
-
( k_2 \cdot k_3 ) 
\,
g^{\alpha \mu }
\Big]
k_1^{\nu } 
+
( k_1 \cdot k_3 ) 
\Big[
k_2^{\nu } g^{\alpha \mu }
-
k_2^{\alpha } g^{\mu \nu }
\Big]
\Big\}
\,
\mathcal{F}_{231}
\\
&&\hspace{-0.25cm}
+ 
\Big\{
\Big[
k_2^{\nu } k_3^{\alpha }
-
( k_2 \cdot k_3 ) 
\,
g^{\alpha \nu }
\Big]
k_1^{\mu } 
+
( k_1 \cdot k_2 )
\Big[
k_3^{\mu } g^{\alpha \nu }
-
k_3^{\alpha } g^{\mu \nu }
\Big]
\Big\}
\,
\mathcal{F}_{312}
\nonumber \\
&&\hspace{-0.25cm}
+ 
\Big[
k_2^{\nu } k_3^{\mu }
-
( k_2 \cdot k_3 ) 
\,
g^{\mu \nu }
\Big]
k_2^{\alpha } 
\,
\mathcal{F}_{232}
+ 
\Big[
k_2^{\nu } k_3^{\mu }
-
( k_2 \cdot k_3 ) 
\,
g^{\mu \nu }
\Big]
k_3^{\alpha } 
\,
\mathcal{F}_{332}.
\nonumber 
\end{eqnarray}
In addition, we derive symmetry relations
among the form factors $\mathcal{F}_{ijl}$
by interchanging $k_2 \leftrightarrow k_3$
and $\mu \leftrightarrow \nu$ within
the Lorentz-invariant amplitude
$\mathcal{M}^{\alpha \mu \nu}$
in Eqs.~(\ref{sym1AMP},~\ref{invAmp}),
leading to the following identities:
\begin{eqnarray}
\mathcal{F}_{211}
&=&
\mathcal{F}_{311}
\,
\big\{
k_{12} \leftrightarrow k_{13}
\big\},
\\
\mathcal{F}_{312} 
&=&
\mathcal{F}_{231} 
\,
\big\{
k_{12} \leftrightarrow k_{13}
\big\},
\\
\mathcal{F}_{212} 
&=&
\mathcal{F}_{331} 
\,
\big\{
k_{12} \leftrightarrow k_{13}
\big\},
\\
\mathcal{F}_{232} 
&=&
\mathcal{F}_{332} 
\,
\big\{
k_{12} \leftrightarrow k_{13}
\big\}.
\end{eqnarray}
Consequently, the one-loop amplitude
can be reformulated into a compact
form as follows:
\begin{eqnarray}
\mathcal{M}^{\alpha \mu \nu} 
&=&
\dfrac{\mathcal{F}_{211}}{k_2 \cdot k_3}
\Big[
( k_2 \cdot k_3 )
\,
k_1^{\nu } 
-
( k_1 \cdot k_3 ) 
\,
k_2^{\nu }
\Big] 
\Big[
k_1^{\mu } k_2^{\alpha }
-
( k_1 \cdot k_2 ) 
\,
g^{\alpha \mu }
\Big]
\nonumber \\
&&\hspace{-0.25cm}
+ 
\mathcal{F}_{231}
\Big\{
\Big[
k_2^{\alpha } k_3^{\mu }
-
( k_2 \cdot k_3 ) 
\,
g^{\alpha \mu }
\Big]
k_1^{\nu } 
+
( k_1 \cdot k_3 ) 
\Big[
k_2^{\nu } g^{\alpha \mu }
-
k_2^{\alpha } g^{\mu \nu }
\Big]
\Big\}
\\
&&\hspace{-0.25cm}
+ 
\mathcal{F}_{232}
\Big[
k_2^{\nu } k_3^{\mu }
-
( k_2 \cdot k_3 ) 
\,
g^{\mu \nu }
\Big]
k_2^{\alpha } 
+
\big(
k_{12} \leftrightarrow k_{13}
,
k_2 \leftrightarrow k_3
,
\mu \leftrightarrow \nu
\big)-\textrm{terms}.
\nonumber 
\end{eqnarray}
The expression for this amplitude
serves as the starting point of our study.
It establishes that three independent form
factors are necessary to compute the decay
rates in the subsequent analysis.
The form factors $\mathcal{F}_{ijl}$,
for $i, j, l = 1, 2, 3$, can be decomposed
into distinct contributions arising from
$\phi$-pole diagrams and one-loop box diagrams
relevant to the decay process, as follows:
\begin{eqnarray}
\mathcal{F}_{i j l}
&=&
\dfrac{i}{\pi^2}
\sum \limits_{\phi \equiv h, H}
\dfrac{
g_{A Z \phi}
}{
k_{23}-M_{\phi}^2
+
i \Gamma_{\phi} M_{\phi} 
}
\Bigg[
g_{\phi \, W^\pm W^\mp}
\times
\mathcal{F}_{i j l}^{\text{Trig}, W^\pm}
\\
&&
+
\sum \limits_{F \equiv f, U_i, D_i}
N_C^F
\,
M_F^2
\times
g_{\phi F \bar{F}} 
\times
\mathcal{F}_{i j l}^{\text{Trig}, F}
+
\sum \limits_{S_Q \equiv H^\pm, H^{\pm\pm}}
g_{\phi S_Q S_Q}
\times
\mathcal{F}_{i j l}^{\text{Trig}, S_Q}
\Bigg]
\nonumber
\\
&&
+
\dfrac{i}{\pi^2}
\sum \limits_{F \equiv f, U_i, D_i}
N_C^F
\,
M_F^2 
\times
g_{A F \bar{F}}
\times 
g^A_{Z F \bar{F}}
\times
(g_{\gamma F \bar{F}})^2
\times
\mathcal{F}_{i j l}^{\text{Box}}.
\nonumber
\end{eqnarray}
Here, $N_C^F$ denotes the number of
colors for the fermion $F$,
taking the value $1$ for leptons and $3$ for quarks.
The coupling is defined as $g_{\gamma F \bar{F}} = e Q_F$,
while the axial component of the coupling
is given by $g^A_{Z F \bar{F}} = -\left(e/s_{2W}\right)\times I_3^F$ for all fermions $F$. Furthermore, the contributions $\mathcal{F}_{ijl}^{\text{Trig}, W^\pm}$, $\mathcal{F}_{ijl}^{\text{Trig}, F}$, and $\mathcal{F}_{ijl}^{\text{Trig}, S_Q}$ originate from
one-loop vertex Feynman diagrams
involving the poles of $\phi \equiv h,~H$. These contributions
arise from loops containing $W^\pm$ bosons, the SM fermions $F \equiv f = e, \mu, \tau, u, c, t, \ldots$, VLFs $F \equiv U_i, D_i$ for $i = 1, 2, \cdots$, and charged scalar Higgs bosons $S_Q \equiv H^\pm,~H^{\pm\pm}$. Additionally, contributions from box diagrams, denoted as $\mathcal{F}_{ijl}^{\text{Box}}$, are incorporated for fermions $F$ in our analysis. The complete set of form factors $\mathcal{F}_{ijl}$ is systematically listed and analytically expressed in the subsequent paragraphs.

Specifically, the form factors $\mathcal{F}_{ijl}^{\text{Trig}, W^\pm}$, evaluated at the poles $\phi$ and arising from loop diagrams with $W^\pm$-boson propagation, can be represented using
the PV-functions as follows:
\begin{eqnarray}
\mathcal{F}_{232}^{\text{Trig}, W^\pm}
&=&
\dfrac{e^2}{2 M_W^2} 
\Big[
4 M_W^2 
C_{0}
-
\big(
M_{\phi}^2
+
6 M_W^2
\big) 
C_{12}
\Big]
(0,k_{23},0,M_W^2,M_W^2,M_W^2),
\\
\mathcal{F}_{211}^{\text{Trig}, W^\pm}
&=&
\mathcal{F}_{231}^{\text{Trig}, W^\pm}
= 0.
\end{eqnarray}
Next, the form factors
$\mathcal{F}_{ijl}^{\text{Trig}, F}$ at
the poles $\phi$, corresponding
to fermion $F$
contributions, are expressed in
the following form:
\begin{eqnarray}
\mathcal{F}_{232}^{\text{Trig}, F}
&=&
(g_{\gamma F \bar{F}})^2  
\times 
\Big[
C_{0}
-
4 C_{12}
\Big]
(0,k_{23},0,M_F^2,M_F^2,M_F^2),
\\
\mathcal{F}_{211}^{\text{Trig}, F}
&=&
\mathcal{F}_{231}^{\text{Trig}, F}
= 0.
\end{eqnarray}
Notably, only the first diagram in Fig.~\ref{FeyDia-HF}
contributes to these form factors.

Moreover, the form factors
$\mathcal{F}_{ijl}^{\text{Trig}, S_Q}$ at
the poles $\phi$, resulting from the contributions
of the charged scalar Higgs $S_Q$, are given by:
\begin{eqnarray}
\mathcal{F}_{232}^{\text{Trig}, S_Q}
&=&
(g_{\gamma S_Q S_Q})^2
\times
C_{12}
(0,k_{23},0,M_{S_Q}^2,M_{S_Q}^2,M_{S_Q}^2),
\\
\mathcal{F}_{211}^{\text{Trig}, S_Q}
&=&
\mathcal{F}_{231}^{\text{Trig}, S_Q}
= 0.
\end{eqnarray}
The coupling is defined as
$g_{\gamma S_Q S_Q} = - e Q_{S_Q}$,
with the charged scalar Higgs masses
given by $M_{S_Q}^2 = M_{H^\pm}^2,~M_{H^{\pm\pm}}^2$,
corresponding to the singly and doubly charged
Higgs bosons. Similarly, the
form factors $\mathcal{F}_{ijl}^{\text{Trig}, S_Q}$
do not receive contributions from the
third diagram in Fig.~\ref{FeyDia-HF}.

Finally, we derive the analytical expressions
for the form factors $\mathcal{F}_{ijl}^{\text{Box}}$,
which originate from fermion $F$ propagating
in one-loop box diagrams. Their explicit
representations are provided below:
\begin{eqnarray}
\mathcal{F}_{232}^{\text{Box}}
&=&
\Big[
D_{0}
+
D_{1}
+
D_{2}
+
3 D_{3}
+
2 D_{13}
+
2 D_{23}
+
2 D_{33}
\Big]
(0,M_Z^2,0,M_A^2;
k_{12},k_{13};\vec{M}_F^2)
\nonumber \\
&&\hspace{0cm}
+\Big[
D_{3}
-
D_{2}
+
2 D_{23}
+
2 D_{33}
\Big]
(M_Z^2,0,0,M_A^2;
k_{12},k_{23}; \vec{M}_F^2)
\nonumber \\
&&
+
\Big[
D_{3}
+ 
2 D_{23}
+ 
2 D_{33}
\Big]
(M_Z^2,0,0,M_A^2;
k_{13},k_{23}; \vec{M}_F^2),
\\
\mathcal{F}_{211}^{\text{Box}}
&=&
\Big[
D_{0}
+
3 D_{2}
+
2 D_{22}
+
4 D_{23}
+
3 D_{3}
+
2 D_{33}
\Big]
(0,M_Z^2,0,M_A^2;
k_{12},k_{13};
\vec{M}_F^2)
\nonumber \\
&&
+
\Big[
D_{0}
+
3 D_{1}
+
2 D_{11}
+
4 D_{12}
+
4 D_{13}
+
3 D_{2}
+
2 D_{22}
\nonumber \\
&&
\hspace{4cm}
+
4 D_{23}
+
3 D_{3}
+
2 D_{33}
\Big]
(M_Z^2,0,0,M_A^2;
k_{12},k_{23}; \vec{M}_F^2 )
\nonumber \\
&&
+
\Big[
D_{0}
+
3 D_{1}
+
2 D_{11}
+
4 D_{12}
+
4 D_{13}
+
3 D_{2}
+
2 D_{22}
\\
&&
\hspace{4cm}
+
4 D_{23}
+
3 D_{3}
+
2 D_{33}
\Big]
(M_Z^2,0,0,M_A^2;
k_{13},k_{23};
\vec{M}_F^2 ),
\nonumber\\
\mathcal{F}_{231}^{\text{Box}}
&=&
\Big[
\dfrac{1}{2} D_{0}
+
D_{2}
+
2 D_{23}
+
3 D_{3}
+
2 D_{33}
\Big]
(0,M_Z^2,0,M_A^2;
k_{12},k_{13};
\vec{M}_F^2)
\nonumber \\
&&
+
\Big[
D_{3}
-
\dfrac{1}{2} D_{0}
-
D_{1}
+
2 D_{13}
-
D_{2}
+
2 D_{23}
+
2 D_{33}
\Big]
(M_Z^2,0,0,M_A^2;
k_{12},k_{23};
\vec{M}_F^2)
\nonumber \\
&&
+
\Big[
\dfrac{1}{2} D_{0}
+
D_{1}
+
2 D_{12}
+
2 D_{13}
+
3 D_{2}
+
2 D_{22}
\nonumber \\
&&
\hspace{3.0cm}
+
4 D_{23}
+
3 D_{3}
+
2 D_{33}
\Big]
(M_Z^2,0,0,M_A^2;
k_{13},k_{23};
\vec{M}_F^2).
\end{eqnarray}
In the above expressions,
the notation
$\vec{M}_F^2 = \{M_F^2, M_F^2, M_F^2, M_F^2\}$
is introduced for notational convenience.

Next, we expand the $1/(d - 4)$-term
in each one-loop integral,
where the dimensional parameter
$d = 4 - 2\epsilon$ is introduced
following the approach of
dimensional regularization.
As a result, the integrals
that do not exhibit divergences
are set to zero when verifying
the $\text{UV}$-finiteness of the calculation.
The $\text{UV}$-divergent parts
of one-loop integrals are expressed
through the dimensionally
regularized divergence parameter
$\Delta =1/\epsilon -\log(4\pi) + \gamma_E$
with Euler-Mascheroni constant
$\gamma_E = 0.5772156\cdots $
as in
Refs.~\cite{Hahn:1998yk, Denner:2016kdg}.
By factorizing all terms involving
$\Delta$ in the above form factors,
we verify that:
\begin{eqnarray}
\mathcal{F}_{232}^{\text{Trig},
W^\pm/F/S_Q}|_{\Delta}
&=& 0.
\end{eqnarray}
Since PV functions, such as $D_j,~D_{ij}, \dots$,
appear in the form factors from one-loop box
diagrams and are free of UV divergences,
it follows that all form factors are
UV-finite.

With all relevant one-loop form
factors determined, we proceed to
compute the partial decay rates
for the process.
Based on the previously derived
expression for the invariant amplitude
$\mathcal{M}^{\alpha \mu \nu}$ corresponding
to the decay channel $A(p) \rightarrow
Z_\alpha (k_1) \gamma_\mu (k_2) \gamma_\nu (k_3)$,
we can directly compute the squared amplitude,
summed over the polarizations of the two photons
and the $Z$ boson. This quantity is essential
for determining the decay rate, which can be
expressed as follows:
\begin{eqnarray}
&&\hspace{-0.7cm}
\sum 
\limits_\text{pol.}
\big|
\mathcal{M}_{A\rightarrow Z\gamma \gamma}
\big|^2
= \nonumber\\
&=&
\mathcal{C}_1
\big|
\mathcal{F}_{211}
\big|^2
+
\hat{\mathcal{C}}_1
\big|
\widetilde{\mathcal{F}}_{211}
\big|^2
+
\mathcal{C}_2
\big|
\mathcal{F}_{231}
\big|^2
+
\hat{\mathcal{C}}_2
\big|
\widetilde{\mathcal{F}}_{231}
\big|^2
+
\mathcal{C}_3
\big|
\mathcal{F}_{232}
\big|^2
+
\hat{\mathcal{C}}_3
\big|
\widetilde{\mathcal{F}}_{232}
\big|^2
\\
&&
+
\mathcal{C}_4
\text{Re}
\big[
\mathcal{F}_{211}
\,
(\widetilde{\mathcal{F}}_{211})^*
\big]
+
\hat{\mathcal{C}}_4
\text{Re}
\big[
\widetilde{\mathcal{F}}_{211}
\,
(\mathcal{F}_{211})^*
\big]
+
\mathcal{C}_5
\text{Re}
\big[
\mathcal{F}_{231}
\,
(\widetilde{\mathcal{F}}_{231})^*
\big]
+
\hat{\mathcal{C}}_5
\text{Re}
\big[
\widetilde{\mathcal{F}}_{231}
\,
(\mathcal{F}_{231})^*
\big]
\nonumber \\
&&
+
\mathcal{C}_6
\text{Re}
\big[
\mathcal{F}_{232}
\,
(\widetilde{\mathcal{F}}_{232})^*
\big]
+
\hat{\mathcal{C}}_6
\text{Re}
\big[
\widetilde{\mathcal{F}}_{232}
\,
(\mathcal{F}_{232})^*
\big]
+
\mathcal{C}_7
\text{Re}
\big[
\mathcal{F}_{211}
\,
(\mathcal{F}_{231})^*
\big]
+
\hat{\mathcal{C}}_7
\text{Re}
\big[
\widetilde{\mathcal{F}}_{211}
\,
(\widetilde{\mathcal{F}}_{231})^*
\big]
\nonumber \\
&&
+
\mathcal{C}_8
\text{Re}
\big[
\mathcal{F}_{231}
\,
(\mathcal{F}_{232})^*
\big]
+
\hat{\mathcal{C}}_8
\text{Re}
\big[
\widetilde{\mathcal{F}}_{231}
\,
(\widetilde{\mathcal{F}}_{232})^*
\big]
+
\mathcal{C}_9
\text{Re}
\big[
\mathcal{F}_{211}
\,
(\widetilde{\mathcal{F}}_{232})^*
\big]
+
\hat{\mathcal{C}}_9
\text{Re}
\big[
\widetilde{\mathcal{F}}_{211}
\,
(\mathcal{F}_{232})^*
\big]
\nonumber \\
&&
+
\mathcal{C}_{10}
\text{Re}
\big[
\mathcal{F}_{232}
\,
(\widetilde{\mathcal{F}}_{231})^*
\big]
+
\hat{\mathcal{C}}_{10}
\text{Re}
\big[
\widetilde{\mathcal{F}}_{232}
\,
(\mathcal{F}_{231})^*
\big]
\nonumber
\end{eqnarray}
where the reduced the factors
$\widetilde{\mathcal{F}}_{i j l}$
are written in terms of the
original ones as given below:
\begin{eqnarray}
\widetilde{\mathcal{F}}_{i j l}
&=&
\mathcal{F}_{i j l}
\,
\big\{
k_{12} \leftrightarrow k_{13}
\big\}
\; \textrm{for}\;
i,j,l =1,2,3.
\end{eqnarray}
The coefficients $\mathcal{C}_i$ for
$i = 1, \ldots, 10$ are defined
in terms of the Lorentz-invariant
kinematic variables $k_{ij}$ and
the corresponding invariant masses,
given by:
\begin{eqnarray}
\mathcal{C}_1
&=&
-\dfrac{
\big(
k_{12}-M_Z^2
\big)^2 
}{2 k_{23}}
\Big[
k_{12}^2
+
M_{A}^2 M_Z^2
+
k_{12} 
\big(
k_{23}-M_{A}^2-M_Z^2
\big)
\Big],
\\
\mathcal{C}_2
&=&
\dfrac{M_Z^2}{8}
\Big[
k_{12}^2
-
M_{A}^2 
\big(
2 k_{12}+4 k_{23} - M_{A}^2
\big)
+
k_{23}
\big(
4 k_{12} + k_{23}
\big)
\Big]
\\
&&
\hspace{0cm}
+
\dfrac{
k_{12} 
}
{
8 M_Z^2
}
\Big[
k_{12}
\big(
k_{12}+k_{23}-M_{A}^2
\big)^2
-
2 M_Z^2 
\big(
k_{12}+2 k_{23}-M_{A}^2
\big) 
\big(
k_{12}+k_{23}-M_{A}^2
\big)
\Big]
,
\nonumber 
\\
\mathcal{C}_3
&=&
\dfrac{k_{23}^2}{8 M_Z^2}
\big(
k_{12}-M_Z^2
\big)^2,
\\
\mathcal{C}_4
&=&
-\dfrac{1}{4 k_{23}}
\Big[
M_{A}^2 M_Z^2
+
k_{12} 
\big(k_{12}
+
k_{23}-M_{A}^2-M_Z^2
\big)
\Big] 
\times
\nonumber \\
&&
\times
\Big[
M_Z^2 
\big(
k_{23}+M_{A}^2
\big)
+
k_{12} 
\big(k_{12}+
k_{23}-M_{A}^2-M_Z^2
\big)
\Big],
\\
\mathcal{C}_5
&=&
\dfrac{1}{8 M_Z^2}
\big(
k_{12}-M_Z^2
\big) 
\big(
k_{12}+k_{23}-M_{A}^2
\big) 
\times
\nonumber \\
&&
\times
\Big[
k_{12}^2
+
M_Z^2 
\big(
k_{23}+M_{A}^2
\big)
+
k_{12} 
\big(
k_{23}-M_{A}^2-M_Z^2
\big)
\Big],
\\
\mathcal{C}_6
&=&
-\dfrac{k_{23}^2}{8 M_Z^2}
\Big[
k_{12}^2
+
M_Z^2 
\big(
k_{23}+M_{A}^2
\big)
+
k_{12} 
\big(
k_{23}-M_{A}^2-M_Z^2
\big)
\Big],
\\
\mathcal{C}_7
&=&
\big(
M_Z^2 - k_{12}
\big) 
\Big[
k_{12}^2
+
M_{A}^2 M_Z^2
+
k_{12} 
\big(
k_{23}-M_{A}^2-M_Z^2
\big)
\Big],
\\
\mathcal{C}_8
&=&
\dfrac{k_{23}^2 }{4} 
\big(
k_{12}-M_Z^2
\big),
\\
\mathcal{C}_9
&=&
\dfrac{k_{23} }{2} 
\Big[
k_{12}^2
+
M_{A}^2 M_Z^2
+
k_{12} 
\big(
k_{23}-M_{A}^2-M_Z^2
\big)
\Big],
\\
\mathcal{C}_{10}
&=&
\dfrac{k_{23}^2 }{4} 
\big(
M_Z^2-k_{12}
\big).
\end{eqnarray}
It is straightforward to
verify that the other
coefficients $\hat{\mathcal{C}}_i$
can be expressed as:
\begin{eqnarray}
\hat{\mathcal{C}}_i
&=&
\mathcal{C}_i
\,
\big|
\{k_{12} \leftrightarrow k_{13}\}.
\end{eqnarray}
Ultimately, the squared amplitude
for the process can be formulated
in a concise manner as:
\begin{eqnarray}
&& \hspace{-0.9cm}
\sum 
\limits_\text{pol.}
\big|\mathcal{M}
_{A\rightarrow Z\gamma \gamma}
\big|^2  
=
\\
&=&
\mathcal{C}_1
\big|
\mathcal{F}_{211}
\big|^2
+
\mathcal{C}_2
\big|
\mathcal{F}_{231}
\big|^2
+
\mathcal{C}_3
\big|
\mathcal{F}_{232}
\big|^2
+
\mathcal{C}_4
\text{Re}
\big[
\mathcal{F}_{211}
\,
(
\widetilde{
\mathcal{F}}_{211}
)^*
\big]
+
\mathcal{C}_5
\text{Re}
\big[
\mathcal{F}_{231}
\,
(\widetilde{\mathcal{F}}_{231})^*
\big]
\nonumber \\
&&
+
\mathcal{C}_6
\text{Re}
\big[
\mathcal{F}_{232}
\,
(\widetilde{\mathcal{F}}_{232})^*
\big]
+
\mathcal{C}_7
\text{Re}
\big[
\mathcal{F}_{211}
\,
(\mathcal{F}_{231})^*
\big]
+
\mathcal{C}_8
\text{Re}
\big[
\mathcal{F}_{231}
\,
(\mathcal{F}_{232})^*
\big]
\nonumber
\\
&&
+
\mathcal{C}_9
\text{Re}
\big[
\mathcal{F}_{211}
\,
(\widetilde{\mathcal{F}}_{232})^*
\big]
+
\mathcal{C}_{10}
\text{Re}
\big[
\mathcal{F}_{232}
\,
(\widetilde{\mathcal{F}}_{231})^*
\big]
+
\big(
k_{12} \leftrightarrow k_{13}
\big).
\nonumber 
\end{eqnarray}
A key advantage of this formulation
is that the decay rates for the process
can be represented using
three independent one-loop form factors.
The relevant form factors, $\mathcal{F}_{211},
~\mathcal{F}_{231}$, and $\mathcal{F}_{232}$,
are systematically compiled in this study,
with their explicit analytic expressions
presented in the above paragraphs. Utilizing
these essential components,
the decay rate of the process is
subsequently evaluated as:
\begin{eqnarray}
\Gamma_{A \rightarrow 
Z \gamma \gamma}
&=&
\dfrac{1}{256 \pi^3 M_{A}^3}
\int \limits_{k_{23}^\text{min}}^{k_{23}^\text{max}}
d k_{23}
\int \limits_{k_{13}^\text{min}}^{k_{13}^\text{max}}
d k_{13}
\,
\sum \limits_\text{pol.}
\big|
\mathcal{M}_{A \rightarrow 
Z \gamma \gamma}
\big|^2.
\end{eqnarray}
The upper and lower integration
limits are generally defined
for the decay process as
\begin{eqnarray}
k_{23}^\text{min}
&=&
0,
\\
\nonumber
k_{23}^\text{max}
&=&
(M_{A} - M_Z)^2,
\\
\nonumber
k_{13}^\text{max(min)}
&=&
\dfrac{1}{2}
\Big\{
M_{A}^2
+
M_Z^2
-
k_{23}
\pm
\Big[
\big(
M_{A}^2
+
M_Z^2
-
k_{23}
\big)^2
-
4 M_{A}^2
M_Z^2
\Big]^{1/2}
\Big\}.
\end{eqnarray}
\section{Applications}
This section presents the
applications of these calculations,
focusing on the phenomenological
implications within the
Two-Higgs-Doublet Model,
including Vector-Like Fermions,
and the Triplet-Higgs Model.
All input parameters from
the SM are adopted as given
in Refs.~\cite{Phan:2024jbx,
Phan:2024vfy, Phan:2024zus}.
\subsection{THDM}
This subsection begins with the computation of the partial decay rates for the considered process, along with their corresponding branching ratios. The numerical results are evaluated at several benchmark points within the allowed parameter space of the THDM,
as summarized in Table~\ref{decayTHDM}. The model includes seven free parameters: $M_h$, $M_H$, $M_A$, $M_{H^{\pm}}$, the mixing angles $t_{\beta}$ and $s_{\beta-\alpha}$, as well as the symmetry-breaking parameter $m_{12}^2 = M^2 s_\beta c_\beta$. For numerical calculations, we set $M_h=125.1$ GeV and $M^2 = M_H^2$. In Table~\ref{decayTHDM}, we adopt $s_{\beta-\alpha} = 0.98$ to satisfy the alignment limit of the SM. As an example, we fix the CP-even Higgs boson mass at $M_H = 400$ GeV and the charged Higgs boson mass at $M_{H^{\pm}} = 200$ GeV. The first column of Table~\ref{decayTHDM} presents variations in $M_A$ and the mixing angle $t_{\beta}$. The second column shows the total decay rates of $A$, computed by taking into account all two-body decay channels of $A$, as described in Ref.~\cite{Aiko:2022gmz}. The third column provides the partial decay rates for the process $A \rightarrow Z\gamma\gamma$. Finally, the last column reports the branching ratios of the decay $A \rightarrow Z\gamma\gamma$.
Our findings indicate that the branching ratios for this decay channel are of the order of $\mathcal{O}(10^{-4})$ at several points within the allowed parameter space. In general, the partial decay rates of $A \rightarrow Z\gamma \gamma$ are inversely proportional to the mixing angle while being directly proportional to the mass of the CP-odd Higgs boson.
Moreover, we introduce VLFs  as $D_1, D_2$ and $U_1, U_2$, with physical masses set to $M_{U_1} = M_{D_1} = 1000$ GeV and $M_{U_2} = M_{D_2} = 1200$ GeV. Their respective electric charges are $Q_U = 5/3$ and $Q_D = 3/2$, while their mixing angles are chosen as $\theta^{U} = \theta^{D} = 0.3$.
With these VLFs contributing to the loop of the considered process,
the resulting partial decay rates and branching ratios are
provided in the second row of the third and last columns of Table~\ref{decayTHDM}.
Additionally, the VLFs in the loop
contribute approximately $10\%$ to the
partial decay rates.
\begin{table}[H]
\begin{center}
\begin{tabular}{
c@{\hspace{1cm}}
c@{\hspace{1cm}}
c@{\hspace{1cm}}
c@{\hspace{1cm}} c}
\hline\hline
($M_{A}$ [GeV] ,~$t_{\beta}$) 
& $\Gamma_{A}^{\textrm{Total}}$ [GeV] 
& $\Gamma_{A \rightarrow 
Z \gamma \gamma}$
[GeV]
& Br\{$A \rightarrow 
Z \gamma \gamma$\}
\\
\hline\hline 
$(200, 2)$
&
$4.636$
&
$(5.641
\pm \, 0.005)
\cdot 10^{-4}$
&
$1.217\cdot 10^{-4}$
\\
&
&
$(6.111 \pm 0.005)\cdot 10^{-4}$
&
$1.318\cdot 10^{-4}$
\\
$(200, 6)$
&
$4.659$
&
$(5.492\pm 0.005)\cdot 10^{-4}$
&
$1.179 \cdot 10^{-4}$
\\
&
&
$(5.872 \pm 0.005)\cdot 10^{-4}$
&
$1.260\cdot 10^{-4}$
\\  \hline
$(400,2)$
&
$16.748$
&
$(2.723 \pm 0.003)
\cdot 10^{-3}$
&
$1.626\cdot 10^{-4}$
\\
&
&
$(2.924 \pm 0.003)
\cdot 
10^{-3}$
&
$1.746\cdot 10^{-4}$
\\
$(400,6)$
&
$14.583$
&
$(2.386
\pm 0.002)
\cdot 
10^{-3}
$
&
$1.636\cdot 10^{-4}$
\\
&
&
$(2.488 \pm 0.002
)\cdot 10^{-3}$
&
$1.706 \cdot 10^{-4}$
\\ \hline
$(600,2)$
&
$109.872$
&
$(1.300 \pm 0.001)
\cdot 10^{-2} $
&
$1.183\cdot 10^{-4}$
\\
&
&
$(1.402 \pm 0.001)
\cdot 10^{-2} $
&
$1.276 \cdot 10^{-4}$
\\
$(600,6)$
&
$104.149$
&
$(1.114\pm 0.001)\cdot 10^{-2}$
&
$1.070\cdot 10^{-4}$
\\
&
&
$(1.176\pm 0.001)
\cdot 10^{-2} $
&
$1.129\cdot 10^{-4} $
\\ \hline
$(800,2)$
&
$345.760$
&
$(3.596\pm 0.001)
\cdot 10^{-2}  $
&
$1.040\cdot 10^{-4}$
\\
&
&
$(3.900 \pm 0.004)
\cdot 10^{-2}  $
&
$1.128\cdot 10^{-4}$
\\
$(800,6)$
&
$337.244$
&
$(2.960\pm 0.003)
\cdot 10^{-2} $
&
$8.774\cdot 10^{-5}$
\\
&
&
$(3.193\pm 0.003)
\cdot 10^{-2}  $
&
$9.468\cdot 10^{-5} $
\\
\hline\hline 
\end{tabular}
\end{center}
\caption{
\label{decayTHDM} 
Partial decay rates and
their corresponding branching ratios
for the considered process at various points within the viable parameter space of the THDM Type II. The partial decay rates for $A \rightarrow Z \gamma \gamma$ are computed using the multidimensional numerical integration package {\tt CUBA}~\cite{Hahn:2004fe}. The first row corresponds to the THDM, while the second row presents results for the THDM with VLF contributions in the loop.
For the above data, 
we have applied a cut on
$k_{23}^{\textrm{min}}
=10^{-2}$ GeV$^2$.
}
\end{table}
Next, we analyze the branching ratios of the decay process $A \rightarrow Z \gamma \gamma$ at a fixed CP-odd Higgs boson mass of $M_A = 600$~GeV, while investigating the impact of varying the charged Higgs mass and the soft-breaking scale parameter $M$. The branching ratio distributions are evaluated for different values of $M$: $M = -200$~GeV (red dotted line), $M = 0$~GeV (blue dotted line), $M = 400$~GeV (green dotted line), and $M = 600$~GeV (black dotted line). The analysis is performed with the CP-even Higgs boson mass fixed at $M_H = 400$~GeV, while the charged Higgs mass is varied within the range $300$~GeV $\leq M_{H^{\pm}} \leq 800$~GeV. The resulting distributions are presented for $t_\beta = 2$ on the left and $t_\beta = 6$ on the right. We observe that the branching ratios predominantly increase within the range $300$~GeV $\leq M_{H^{\pm}} \leq 500$~GeV and remain nearly constant for $500$~GeV $\leq M_{H^{\pm}} \leq 700$~GeV. However, they decrease sharply for $M_{H^{\pm}} \geq 700$~GeV. Additionally, the branching ratios exhibit slight variations with changes in the soft-breaking scale parameter $M$.
Beyond $M_{H^{\pm}} \geq 700$~GeV, the results show that the branching ratios change insignificantly with respect to $M$, as the contribution of the charged Higgs boson in the loop becomes negligible at higher masses.
\begin{figure}[H]
\centering
\begin{tabular}{cc}
\hspace{-4cm}
Br$\{ A \rightarrow Z \gamma \gamma\}$
$\times 10^{-4}$
&
\hspace{-4cm}
Br$\{ A \rightarrow Z \gamma \gamma\}$
$\times 10^{-4}$
\\
\includegraphics[width=8cm, height=6cm]
 {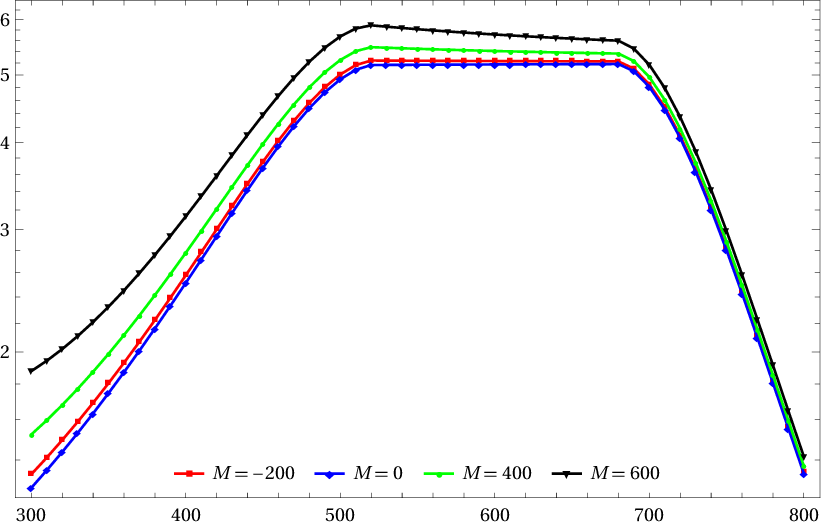}
 &
\includegraphics[width=8cm, height=6cm]
 {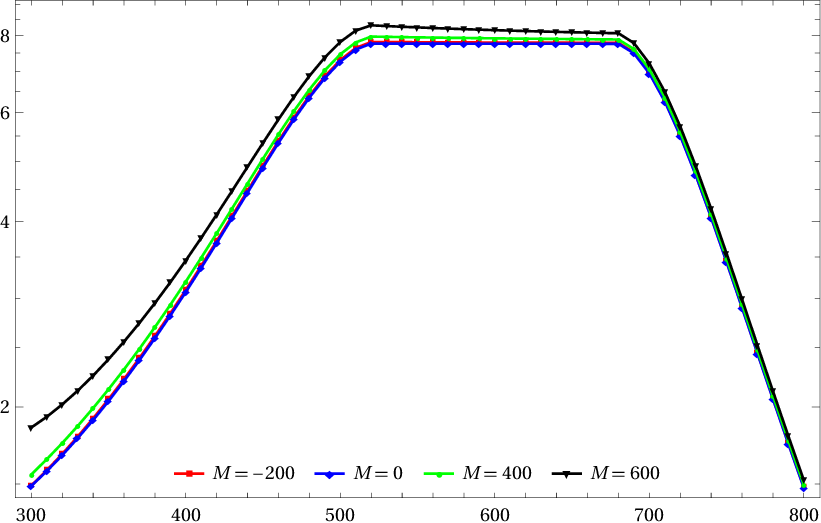}
  \\
\hspace{6cm}
$M_{H^\pm}$ [GeV]
&
\hspace{6cm}
$M_{H^\pm}$ [GeV]
 \end{tabular}
\caption{\label{MATHDM}  
The branching ratios for the process $A \rightarrow Z \gamma \gamma$ are examined at $M_A = 600$~GeV. The distributions are analyzed for various values of the soft-breaking scale parameter: $M = -200$~GeV (red dotted line), $M = 0$~GeV (blue dotted line), $M = 400$~GeV (green dotted line), and $M = 600$~GeV (black dotted line). The study is performed with the CP-even Higgs mass fixed at $M_H = 400$~GeV, while the charged Higgs mass is varied within the range $300$~GeV~$\leq M_{H^{\pm}} \leq 800$~GeV. The resulting distributions are depicted for $t_\beta = 2$ on the left and $t_\beta = 6$ on the right.
}
\end{figure}
We now focus on the branching ratios of the decay process
$A \rightarrow Z \gamma \gamma$ with the inclusion of VLFs in the loop for $M_A = 600$~GeV. In this analysis, the distributions are examined as functions of the degenerate mass splitting of VLFs, denoted as $\Delta M_F$, where $\Delta M_F = M_{U_2} - M_{U_1} = M_{D_2} - M_{D_1}$, as well as the parameter $t_\beta$. Specifically, we consider the following scenarios: $\Delta M_F = 50$~GeV with $t_\beta = 2$ (red dotted line), $\Delta M_F = 50$~GeV with $t_\beta = 6$ (blue dotted line), $\Delta M_F = 200$~GeV with $t_\beta = 2$ (green dotted line), and $\Delta M_F = 200$~GeV with $t_\beta = 6$ (magenta dotted line). The mixing angles are fixed at $\theta^{U} = \theta^{D} = 0.2$. Furthermore, we extend our analysis by varying the charged Higgs mass within the range $300$~GeV$\leq M_{H^{\pm}} \leq 800$~GeV. The remaining physical masses of the VLFs are set to $M_F = M_{U_1} = M_{D_1}$ and $M_{U_2} = M_{D_2} = 1200$~GeV. Additionally, the CP-even Higgs mass is chosen
as in the previous case, or $M_H = 400$~GeV.
The results indicate that the branching ratios depend on the charged Higgs mass, exhibiting behaviors similar to those observed in previous distributions. While they
show only slight variations with the degenerate mass splitting of VLFs,
it indicates a substantial dependence on the mixing angle $t_\beta$. This behavior is expected, as the coupling of VLFs to the CP-even Higgs bosons is governed by the mixing angle $t_\beta$.
\begin{figure}[H]
\centering
\hspace{-10cm}
Br$\{ A \rightarrow Z \gamma \gamma\}$
$\times 10^{-4}$
\\
\includegraphics[width=14cm, height=8cm]
 {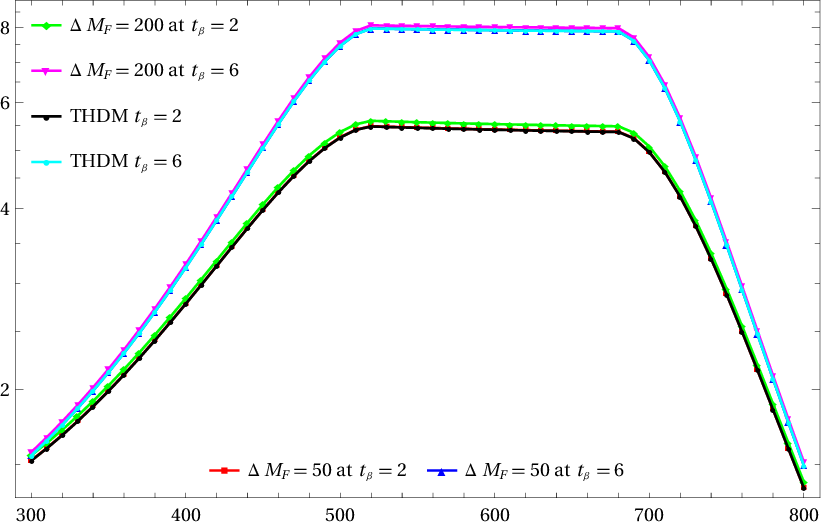}
 \\
\hspace{12cm}
$M_{H^\pm}$ [GeV]
\caption{\label{MATHDM}  
We analyze the branching ratios of the process
$A \rightarrow Z \gamma \gamma$ in the presence
of VLFs in the loop for $M_A = 600$~GeV.
The distributions are evaluated as functions
of the degenerate mass splitting $\Delta M_F$,
defined as $\Delta M_F = M_{U_2} - M_{U_1}
= M_{D_2} - M_{D_1}$, and the mixing
parameter $t_\beta$.  Additionally,
we examine how variations in the
charged Higgs mass, within the
range $300$~GeV $\leq M_{H^{\pm}} \leq 800$~GeV,
affect the branching ratios.}
\end{figure}
\subsection{THM} 
In the numerical analysis for the THM, 
we consider the limiting case where $v_{\Delta} \rightarrow 0$.
In this regime, the parameter $t_{\beta^{\pm}}$ is defined as
$t_{\beta^{\pm}} = \frac{\sqrt{2} v_{\Delta}}{v_{\Phi}} =
\frac{\sqrt{2} v_{\Delta}}{\sqrt{v^2 - 2v_{\Delta}^2}}$
which leads to $t_{\beta^{0}} = \sqrt{2} t_{\beta^{\pm}}\rightarrow 0$. The mixing angle is set to $c_{\alpha} = 0.95$ and the mass of the doubly charged Higgs boson is fixed at $M_{H^{\pm\pm}} = 300$~GeV, while the CP-even Higgs boson has a mass of $M_H = 400$~GeV. The mass of the singly charged Higgs boson is determined through the following relation
(see Refs.~\cite{Aoki:2012jj,Hue:2023tdz}):
\begin{eqnarray}
\label{MSQ}
M_{H^{\pm}}^2
= \frac{ M_A^2 + M_{H^{\pm\pm}}^2 }{2}.
\end{eqnarray}
Table~\ref{decayTHM} presents the partial decay rates for the process $A \to Z \gamma \gamma$ along with the corresponding branching ratios, evaluated for various values of $M_A$ and $\lambda_1$. Our findings indicate that the partial decay rates scale proportionally with $M_A$. Besides that, for a fixed value of $M_A$, the partial decay rates of the process $A \to Z \gamma \gamma$ are found to be proportional to $\lambda_1$.
In the considered limit, the three dominant two-body decay channels of the CP-odd Higgs boson $A$ are $A\rightarrow W^{\pm}H^{\mp}$, $A\rightarrow Zh$, and $A\rightarrow ZH$. The decay rate of $A\rightarrow W^{\pm}H^{\mp}$ exhibits a dependence on the charged Higgs mass. In the mass range $400~\text{GeV} \leq M_A \leq 600~\text{GeV}$, Eq.~(\ref{MSQ}) implies that $M_A \sim M_{H^{\pm}}$. Consequently, the decay rates of $A\rightarrow W^{\pm}H^{\mp}$ are suppressed in this region, leading to a reduction in the total decay width compared to other mass ranges. As a result, the branching ratio of $A \rightarrow Z \gamma \gamma$ is enhanced and can reach $\mathcal{O}(10^{-2})$ in this region.
\begin{table}[H]
\begin{center}
\begin{tabular}{
c@{\hspace{1cm}}
c@{\hspace{1cm}}
c@{\hspace{1cm}}
c@{\hspace{1cm}}}
\hline\hline
($M_{A}$~[GeV], $\lambda_1 $)
& $\Gamma_{A}^{\textrm{Total}}$
[GeV]
& $\Gamma_{A \rightarrow 
Z \gamma \gamma}$
[GeV]
& Br($A \rightarrow 
Z \gamma \gamma$)
\\
\hline 
$(200, -2)$
&
$174.149$
&
$(5.097 \pm 0.005)\cdot 10^{-3}$
&
$2.927 \cdot 10^{-5}$
\\
$(200,~~0)$
&
$174.149$
&
$(6.995\pm 0.007 )\cdot 10^{-3}$
&
$4.017 \cdot 10^{-5}$
\\
$(200, +2)$
&
$174.149$
&
$(9.599 \pm 0.009 )\cdot 10^{-3}$
&
$5.512\cdot 10^{-5}$
\\ \hline
$(400, -2)$
&
$4.822$
&
$(4.168\pm 0.004)\cdot 10^{-2}$
&
$8.644 \cdot 10^{-3}$
\\
$(400,~~0)$
&
$4.822$
&
$(5.215 \pm  0.005)\cdot 10^{-2}$
&
$1.081 \cdot 10^{-2}$
\\
$(400, +2)$
&
$4.822$
&
$(6.377 \pm 0.006)\cdot 10^{-2}$
&
$1.322 \cdot 10^{-2}$
\\
\hline
$(600, -2)$
&
$59.597$
&
$(2.016\pm 0.002)\cdot 10^{-1}$
&
$3.383 \cdot 10^{-3}$
\\
$(600,~~0)$
&
$59.597$
&
$(2.478 \pm 0.002)\cdot 10^{-1}$
&
$4.157 \cdot 10^{-3}$
\\
$(600, +2)$
&
$59.597$
&
$(2.987 \pm 0.002)\cdot 10^{-1} $
&
$5.011 \cdot 10^{-3}$
\\
\hline
$(800, -2)$
&
$332.645$
&
$(5.418\pm 0.005)\cdot 10^{-1}$
&
$1.629\cdot 10^{-3}$
\\ 
$(800,~~0)$
&
$332.645$
&
$(6.597 \pm 0.007)\cdot 10^{-1}$
&
$1.983\cdot  10^{-3}$
\\ 
$(800, +2)$
&
$332.645$
&
$(7.899\pm 0.008)\cdot 10^{-1}$
&
$2.375\cdot 10^{-3}$
\\ 
\hline
\hline
\end{tabular}
\end{center}
\caption{
\label{decayTHM}  
Partial decay rates for the considered
process at several points within the
viable parameter space of the THM.
The partial decay rates for 
$A \rightarrow Z \gamma \gamma$
are generated by using the 
multidimensional numerical 
integration {\tt CUBA} 
program~\cite{Hahn:2004fe}. For these
calculations, we have applied a cut on
$k_{23}^{\textrm{min}}=10^{-2}$ GeV$^2$.
}
\end{table}
Branching ratios for the process
$A \rightarrow Z \gamma \gamma$ are analyzed at $M_A = 600$~GeV. The distributions are evaluated for considering different values of
the doubly charged Higgs boson mass over the range
$300$~GeV~$\leq M_{H^{\pm\pm}} \leq 800$~GeV. Additionally, we consider the limit $v_{\Delta} \to 0$, adopting $v_{\Delta} = 1$~MeV as a representative case.
The mass of the CP-even Higgs boson is set to
$M_H = 400$~GeV.
The singly charged Higgs boson mass is  determined by
Eq.~(\ref{MSQ}). The distributions are presented for various values of $\lambda_1$, with $\lambda_1 = -4$ (red dotted line), $\lambda_1 = -2$ (green dotted line), $\lambda_1 = 0$ (blue dotted line), $\lambda_1 = +2$ (magenta dotted line), and $\lambda_1 = +4$ (black dotted line). It is observed that the branching ratios are proportional to the values of $\lambda_1$, with this dependence being more prominent in the low-mass region of the doubly charged Higgs and less significant at higher values of $M_{H^{\pm\pm}}$. This behavior can be attributed to the contribution of charged Higgs bosons in the loop, which has a greater impact on the partial decay rates of the process $A \rightarrow Z \gamma \gamma$ in the low-mass regime of
$M_{H^{\pm\pm}}$. Notably, for a fixed value of
$\lambda_1$, the dependence of the branching ratios on the doubly charged Higgs mass varies depending on the sign of $\lambda_1$.
Specifically, the branching ratios increase with $M_{H^{\pm\pm}}$ for $\lambda_1 \leq 0$, whereas they decrease with $M_{H^{\pm\pm}}$ for $\lambda_1 \geq 0$.
\begin{figure}[H]
\centering
\hspace{-10cm}
Br$\{ A \rightarrow Z \gamma \gamma\}$
$\times 10^{-3}$
\\
\includegraphics[width=14cm, height=8cm]
{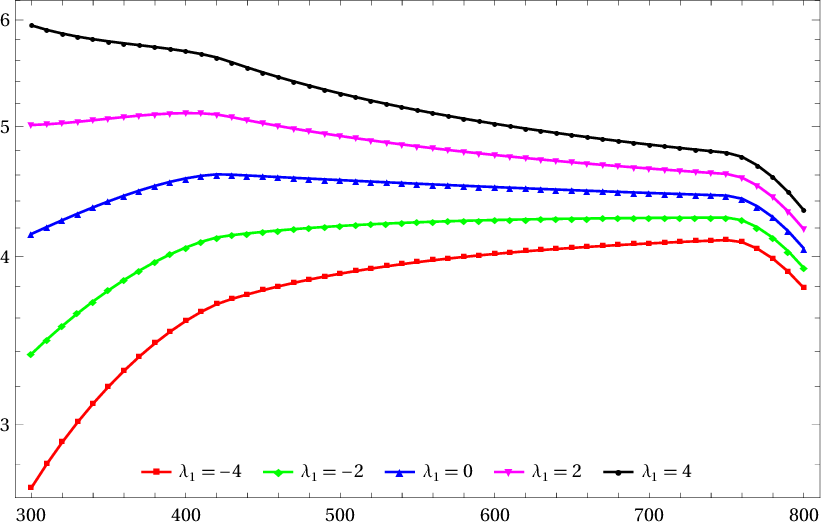}
\\
\hspace{12cm}
$M_{H^{\pm\pm}}$ [GeV]
\caption{\label{MATHDM}  
The branching ratios for the process
$A \rightarrow Z \gamma \gamma$ are
studied at $M_A = 600$~GeV. The distributions are
analyzed across a range of doubly charged Higgs boson
masses, specifically within $300$~GeV~$\leq M_{H^{\pm\pm}} \leq 800$~GeV. The differential distributions are illustrated for
different values of $\lambda_1$, specifically:
$\lambda_1 = -4$ (red dotted line), $\lambda_1 = -2$
(green dotted line), $\lambda_1 = 0$ (blue dotted line),
$\lambda_1 = +2$ (magenta dotted line), and
$\lambda_1 = +4$ (black dotted line).
}
\end{figure}
\section{Conclusions} 
A general one-loop contribution to the rare decay process $A \rightarrow Z \gamma \gamma$ has been evaluated within the frameworks of the Two-Higgs-Doublet Model (including vector-like fermions) and the Triplet-Higgs Model. In our phenomenological investigation, we explored the branching ratios of this decay channel in the viable parameter space of these models. Our findings indicate that the branching ratio can reach
$\mathcal{O}(10^{-4})$ in the Two-Higgs-Doublet Model and $\mathcal{O}(10^{-2})$ in the Triplet-Higgs Model at specific parameter points. Furthermore, the presence of vector-like fermions in the loop leads to a notable shift in the partial decay rates, modifying them by approximately $10\%$. Additionally, we have examined the dependence of the branching ratios on key theoretical parameters, including the charged Higgs mass, mixing angles, and the soft-breaking scale, providing deeper insights into the phenomenological implications of these Higgs-extended frameworks. The calculations presented in this work can be extended and applied to other BSM scenarios.
\\

\noindent
{\bf Acknowledgment:}~
This research is funded by Vietnam
National Foundation for Science and
Technology Development (NAFOSTED) under
the grant number $103.01$-$2023.16$.
\section*{Appendix A: Two Higgs      
Doublet Models}                      
The first example of the Standard Model Higgs extensions
considered in this work is the THDM, summarized in this appendix
based on Ref.~\cite{Branco:2011iw}. A comprehensive
review of the theoretical and phenomenological
aspects of the THDM can be found in
Ref.~\cite{Branco:2011iw}. The scalar
potential, which satisfies renormalization
conditions and preserves gauge
invariance, is given by:
\begin{eqnarray} 
\label{THDMVLF-Pot}
\mathcal{V}^{\textrm{THDM}}
(\Phi_1, \Phi_2) &=&
\sum\limits_{j=1}^2
m^2_{jj}\Phi_j^\dagger\Phi_j
-
\left(m^2_{12}\Phi_1^\dagger\Phi_2
+
{\rm H.c.}\right)
+
\frac{1}{2}
\sum\limits_{j=1}^2
\lambda_j\left(\Phi_j^\dagger\Phi_j\right)^2
\nonumber \\
&&
+\lambda_3\Phi_1^\dagger\Phi_1\Phi_2^\dagger\Phi_2
+\lambda_4\Phi_1^\dagger\Phi_2\Phi_2^\dagger\Phi_1
+\left[\frac{1}{2}
\lambda_5\left(\Phi_1^\dagger\Phi_2\right)^2
+{\rm H.c.}\right].
\end{eqnarray}
Due to the stringent constraints on flavor-changing
neutral currents (FCNCs) at tree level imposed by
experimental data, the $Z_2$ symmetry, defined as
$\Phi_1 \rightarrow \Phi_1$,
$\Phi_2 \rightarrow -\Phi_2$, is applied to
the scalar potential.
This symmetry is softly broken by the term
$ m^2_{12} \Phi_1^\dagger \Phi_2 + {\rm H.c.} $
in the Lagrangian, where the soft-breaking
scalar parameter $m^2_{12}$ is explicitly incorporated.
In the present study, we assume a CP-conserving scenario,
meaning that all bare parameters in the
potential are considered to be real.
Under electroweak symmetry breaking (EWSB),
the two scalar doublets in the scalar potential
can be parameterized around their vacuum
expectation values (VEVs) $(v_1,~v_2)$ as follows:
\begin{eqnarray}
\label{representa-htm}
\Phi_k &=&
\begin{bmatrix}
\phi_k^+ \\
(v_k
+
\phi^0_k
+
i
\varphi^0_k)
/\sqrt{2}
\end{bmatrix}
\quad \textrm{for}
\quad k=1,2.
\end{eqnarray}
The vacuum expectation value (VEV) is subsequently calculated as $v=\sqrt{v_1^2+v_2^2} \approx 246$ GeV to be consistent with the SM.
After EWSB, the physical content of the THDM consists of the following scalar particles: two CP-even Higgs bosons\textemdash{}one corresponding to the discovered SM-like Higgs at the LHC ($h$) and the other a heavier Higgs ($H$), a CP-odd Higgs ($A$), and a pair of singly charged Higgs bosons ($H^{\pm}$). The masses of these physical scalars are determined by diagonalizing the corresponding mass matrices in their flavor bases. The explicit forms of the rotation matrices are given below:
\begin{eqnarray}
\begin{pmatrix}
\phi^0_1\\
\phi^0_2
\end{pmatrix}
&=&
\begin{pmatrix}
c_{\alpha}
&
-s_{\alpha}
\\
s_{\alpha}
&
c_{\alpha}
\end{pmatrix}
\begin{pmatrix}
H\\
h
\end{pmatrix},
\\
\begin{pmatrix}
\phi_1^{\pm}\\
\phi_2^{\pm}
\end{pmatrix}
&=&
\begin{pmatrix}
c_{\beta}
&
-s_{\beta}
\\
s_{\beta}
&
c_{\beta}
\end{pmatrix}
\begin{pmatrix}
G^{\pm}\\
H^{\pm}
\end{pmatrix},
\\
\begin{pmatrix}
\varphi^0_1\\
\varphi^0_2
\end{pmatrix}
&=&
\begin{pmatrix}
c_{\beta}
&
-s_{\beta}
\\
s_{\beta}
&
c_{\beta}
\end{pmatrix}
\begin{pmatrix}
G^{0}\\
A
\end{pmatrix}.
\end{eqnarray}
We employ the simplified notations $s_{\theta} = \sin\theta$, $c_{\theta} = \cos\theta$, and $t_{\theta} = \tan\theta$. The mixing angle $\beta$ is defined as the ratio of the two vacuum expectation values, given by $t_{\beta} = v_2/v_1$. In this framework, the unphysical bosons\textemdash{}namely, the neutral Goldstone boson ($G^0$) and the charged Goldstone bosons ($G^{\pm}$) \textemdash{}play a crucial role in generating masses for
the neutral gauge boson $Z$ and the charged vector bosons $W^{\pm}$, respectively. The remaining physical scalar masses are given by:
\begin{eqnarray}
M_{h}^{2} &=& M_{11}^{2}
s_{\beta-\alpha}^{2}
+ M_{22}^{2}c_{\beta-\alpha}^{2}
+M_{12}^{2}s_{2(\beta-\alpha)},
\\
M_{H}^{2} &=&
M_{11}^{2}c_{\beta-\alpha}^{2}
+M_{22}^{2}s_{\beta-\alpha}^{2}
-M_{12}^{2}s_{2(\beta-\alpha)},
\\
M_{A}^{2}  &=&
M^{2}-\lambda_{5}v^{2},
\\
M_{H^{\pm}}^{2}
&=&
M^{2}-
\frac{1}{2}
\lambda_{45}
v^{2}.
\end{eqnarray}
The parameter $M^2$
is defined as
$M^{2}=m_{12}^{2}/
(s_{\beta}c_{\beta})$
and $M_{ij}$
for $i,j =1,2$ are given
by
\begin{eqnarray}
M_{11}^{2}&=&
(\lambda_{1}c_{\beta}^{4}
+\lambda_{2}s_{\beta}^{4})v^{2}
+\frac{v^{2}}{2}\;
\lambda_{345}
\; s_{2\beta}^{2},
\\
M_{12}^{2} &=&
M_{21}^{2}  =
-\frac{v^{2}}{2}
\Big[
\lambda_{1}c_{\beta}^{2}
-\lambda_{2}s_{\beta}^{2}
-
\lambda_{345}\; c_{2\beta}
\Big]
s_{2\beta},
\\
M_{22}^{2}
&=&
M^{2}
+ \frac{v^{2}}{4}
\Big[
\lambda_{12}
-2\lambda_{345}
\Big]
s_{2\beta}^{2}.
\end{eqnarray}
Here, the shorthand notation
$\lambda_{ij\cdots} = \lambda_{i}
+ \lambda_{j} + \cdots$ is adopted
for compactness.

Finally, we examine the Yukawa sector in THDM.
It is well-established that the introduction of a discrete $Z_2$-symmetry in the THDM is crucial to suppressing tree-level flavor-changing neutral currents. In this context, the Yukawa interactions in the various THDM types are categorized based on the $Z_2$-symmetry. These types are labeled as Type-I, Type-II, Type-X, and Type-Y, with their corresponding $Z_2$-charge assignments summarized in Table~\ref{Z2-assignment}~\cite{Aoki:2009ha}. Consequently, the Yukawa interactions can be expressed in the following general form:
\begin{eqnarray}
{\mathcal L}^{\textrm{THDM}}_\text{Y}
&=&-\sum_{f=u,d,\ell}
\left(
g_{hff}^{\textrm{THDM}}
h{\overline f}f
+g_{Hff}^{\textrm{THDM}} H{\overline
f}f
-i
g_{Aff}^{\textrm{THDM}} A {\overline f}
\gamma_5f
\right)
+\cdots
\\
&=&-\sum_{f=u,d,\ell} 
\left( \frac{m_f}{\sqrt{2}v}\xi_h^f{\overline
f}fh
+\frac{m_f}{\sqrt{2}v}\xi_H^f{\overline
f}fH
-i\frac{m_f}{v}\xi_A^f{\overline f}
\gamma_5fA
\right)
+\cdots
.
\end{eqnarray}
The projection operators $P_{L(R)}$
are used to denote the left- (right-)
handed components of fermions.
The elements $V_{ud}$ of the
Cabibbo-Kobayashi-Maskawa (CKM) matrix,
which describe the quark mixing,
also appear in the Yukawa Lagrangian.
\begin{table}[H]
\begin{center}
\begin{tabular}{c|ccccccc|cccccc}
\hline\hline
Types & $\Phi_{1}$ & $\Phi_{2}$ &$Q_L$
&$L_L$&$u_R$&$d_R$&$e_R$&$\xi^u_A$ 
&$\xi^d_A$&$\xi^{\ell}_A$ 
&$\xi_h^u$
&$\xi_h^d $
&$\xi_h^{\ell}$
\\\hline\hline
I & $+$ & $-$ 
& $+$ & $+$ & $-$ 
& $-$ & $-$ & 
$ct_\beta$ 
& $-ct_\beta$ 
& $ct_\beta$ 
&
$\dfrac{c_\alpha}{
s_\beta}$
&
$
\dfrac{c_\alpha}{
s_\beta}$
&
$
\dfrac{c_\alpha}{
s_\beta}$
\\ \hline
II & $+$ & $-$ & $+$ & $+$
& $-$ & $+$ & $+$ &
$-ct_\beta$
&$-t_\beta$
&$-t_\beta$ 
&
$
\dfrac{c_\alpha}{
s_\beta}$
&
$-
\dfrac{s_\alpha}{
c_\beta}$
&
$-
\dfrac{s_\alpha}{
c_\beta}
$
\\ \hline
X & $+$ & $-$ & $+$
& $+$ & $-$ & $-$ & $+$
& 
$-ct\beta$ 
& $ct_\beta$
& $-t_\beta$ 
&
$
\dfrac{c_\alpha}
{s_\beta}$
&
$
\dfrac{c_\alpha}{
s_\beta}$
&
$-
\dfrac{s_\alpha}
{c_\beta}$
\\ \hline
Y & $+$ & $-$
& $+$ & $+$ & $-$
& $+$ & $-$ &
$-ct_\beta$
&$-t_\beta$
&$ct_\beta$ 
&
$
\dfrac{c_\alpha}
{s_\beta}$
&
$
-
\dfrac{s_\alpha}
{c_\beta}$
&
$
\dfrac{c_\alpha}
{s_\beta}$
\\
\hline
\hline
\end{tabular}
\caption{
\label{Z2-assignment}
The $Z_2$ charge assignments and the factors $\xi^f_{A(h)}$ ($f = u, d, \ell$) corresponding to the four THDM types are provided. The Yukawa couplings of the CP-even Higgs $H$ to a pair of fermions, denoted as $\xi^f_H$, are obtained by exchanging $c_\alpha$ and $s_\alpha$ in $\xi^f_h$. Furthermore, we adopt the notation $c t_{\beta} = \cot \beta$ and $t_{\beta} = \tan \beta$.
}
\end{center}
\end{table}
The couplings involved in the
calculations of this process within the
THDM framework are available in
Refs.~\cite{Hue:2023tdz, Phan:2024vfy, Phan:2024jbx}.
\section*{Appendix B: THDM-II with
Vector-Like Fermions}
We further consider the Type-II
Two-Higgs-Doublet Model with Vector-Like
Fermions (THDM-II-VLF) in this study. This review follows Ref.~\cite{Song:2019aav}. Specifically, the Standard Model fermion sector is extended by introducing VLFs in both the $SU(2)$ doublet and singlet representations, incorporating them as additional degrees of freedom in the theory, given by:
\begin{eqnarray}
\mathcal{Q}_{L}
=
\left(\begin{array}{c}
U'_L  \\
D'_L 
\end{array}
\right), \; 
\mathcal{Q}_{R}
=
\left(
\begin{array}{c}
U'_R  \\
D'_R 
\end{array}
\right), 
\; U_L,~U_R,~D_L,~D_R.
\end{eqnarray}
Here, the notations $U$ and $D$ represent up-type and down-type fermions in the doublets, respectively. In this framework, we consider different types of VLFs. Specifically, $(X, T)$ denotes vector-like quarks (VLQs) with electric charges
$\left(\frac{5}{3},\frac{2}{3}\right)$; $(T, B)$ corresponds to VLQs with electric charges $\left( \frac{2}{3}, -\frac{1}{3} \right)$; $(B, Y)$ represents VLQs with electric charges $\left( -\frac{1}{3}, -\frac{4}{3} \right)$; and $(N, E)$ are vector-like leptons (VLLs) with electric charges $(0, -1)$. The mixing matrix that diagonalizes the mass terms and yields the physical VLFs is given by:
\begin{eqnarray}
\left(
\begin{array}{cc}
D'  \\
D 
\end{array}
\right)
= 
\left(
\begin{array}{cc}
c_{\theta}^D & s_{\theta}^D \\
-s_{\theta}^D & c_{\theta}^D
\end{array}\right)
\left(
\begin{array}{cc}
D_1  \\
D_2 
\end{array}
\right)
,
\quad 
\left(
\begin{array}{cc}
U'  \\
U 
\end{array}
\right)
=
\left(
\begin{array}{cc}
c_{\theta}^U & s_{\theta}^U \\
-s_{\theta}^U & c_{\theta}^U
\end{array}
\right)
\left(
\begin{array}{cc}
U_1  \\
U_2 
\end{array}
\right).
\label{VLFmixing}
\end{eqnarray}
The Lagrangian for the mass
and Yukawa interactions of the
THDM-II-VLF is then written as
follows:
\begin{eqnarray}
-\mathcal{L}^{\textrm{VLF}}_{Y}&=&
\Big[
Y_D\bar{\mathcal{Q}}\Phi_1D
+Y_U\bar{\mathcal{Q}}\tilde{\Phi}_2U
+H.c.
\Big]
+M_{\mathcal{Q}}
\bar{\mathcal{Q}}
\mathcal{Q}
+M_{U}\bar{U}{U}
+M_{D}\bar{D}{D}
\end{eqnarray}
The notation $\tilde{\Phi} = i\tau_2 \Phi^*$ and the relations $Y_U^L = Y_U^R = Y_U$, $Y_D^L = Y_D^R = Y_D$ are employed in the above equations. The mass matrices in the flavor basis for the fields $(D', D)$ and $(U', U)$ are provided below:
\begin{eqnarray}
\mathcal{M}_D=
\left(
\begin{array}{cc}
M_{D'} 
& 
\frac{1}{\sqrt{2}}Y_Dvc_{\beta} 
\\
\frac{1}{\sqrt{2}}Y_Dvc_{\beta} 
& 
M_D
\end{array}
\right),
\quad
\mathcal{M}_U=
\left(
\begin{array}{cc}
M_{U'} 
& 
\frac{1}{\sqrt{2}}Y_U
\; v \; s_{\beta} 
\\
\frac{1}{\sqrt{2}}Y_U
\; v \; s_{\beta} 
& 
M_U
\end{array}
\right)
\end{eqnarray}
The mass matrices of the VLFs are diagonalized
using the rotation matrices given in
Eq.~\eqref{VLFmixing}. The physical masses
of the VLFs are subsequently determined as
\begin{eqnarray}
\label{massD1D2}
M_{D_1}
&=&
(c_{\theta}^D)^2M_{D'}
+(s_{\theta}^D)^2M_D
-\sqrt{2}vc_{\beta}
s_{\theta}^Dc_{\theta}^DY_D
,
\\
M_{D_2}
&=&
(s_{\theta}^D)^2M_{D'}
+(c_{\theta}^D)^2M_D
+\sqrt{2}vc_{\beta}
s_{\theta}^Dc_{\theta}^DY_D.
\end{eqnarray}
The mass for up-types of VLFs are given 
\begin{eqnarray}
\label{massU1U2}
M_{U_1}
&=&
(c_{\theta}^U)^2M_{U'}
+(s_{\theta}^U)^2M_U
-\sqrt{2}vs_{\beta}
s_{\theta}^Uc_{\theta}^UY_U 
,
\\
M_{U_2}
&=&
(s_{\theta}^U)^2M_{U'}
+(c_{\theta}^U)^2M_U
+\sqrt{2}vs_{\beta}
s_{\theta}^Uc_{\theta}^UY_U. 
\end{eqnarray}
The VLFs mixing angles are computed below:
\begin{eqnarray}
\label{thetaUD}
t_{2\theta}^{D}&=&
\frac{\sqrt{2}Y_D\; v\; c_{\beta}}{M_{D'}-M_D}, 
\quad 
t_{2\theta}^{U} =
\frac{\sqrt{2}Y_U\;
v\; s_{\beta}}{M_{U'}-M_U}.
\end{eqnarray}
The Yukawa couplings can be formulated
in terms of the mixing angles and physical
masses of the VLFs in the following manner:
\begin{eqnarray}
 \label{YUD}
 Y_U &=&\dfrac{(M_{U_1}-M_{U_2})s_{2\theta}^U }
 {\sqrt{2}\; v\; s_{\beta} \; c_{4\theta}^U}, \\
 Y_D &=&\dfrac{(M_{D_1}-M_{D_2})s_{2\theta}^D }
 {\sqrt{2}\; v\; c_{\beta} \; c_{4\theta}^D}.
\end{eqnarray}
We can obtain all the
couplings from the Yukawa
Lagrangian in the following way:
\begin{eqnarray}
-\mathcal{L}^{\textrm{VLF}}_Y(\phi_j U\bar{U})
&=&
-\sqrt{2}Y_U s_{\theta}^Uc_{\theta}^Uc_{\alpha}h\bar{U}_{\text{1}}U_{\text{1}}
+\sqrt{2}Y_U s_{\theta}^Uc_{\theta}^Uc_{\alpha}h\bar{U}_{\text{2}}U_{\text{2}}
+\frac{Y_U[(c_{\theta}^U)^2-(s_{\theta}^U)^2]c_{\alpha}}{\sqrt{2}}h\bar{U}_{2}U_{1}
\nonumber\\
&&
+\frac{Y_U[(c_{\theta}^U)^2
-(s_{\theta}^U)^2]c_{\alpha}}
{\sqrt{2}}h\bar{U}_{1}U_{2}
-\sqrt{2}Y_U s_{\theta}^Uc_{\theta}^Us_{\alpha}
H\bar{U}_{\text{1}}U_{\text{1}}
+\sqrt{2}Y_U s_{\theta}^Uc_{\theta}^Us_{\alpha}
H\bar{U}_{\text{2}}U_{\text{2}} 
\nonumber\\
&&
+\frac{Y_U[(c_{\theta}^U)^2
-(s_{\theta}^U)^2]s_{\alpha}}{\sqrt{2}}
H\bar{U}_{2}U_{1}
+\frac{Y_U[(c_{\theta}^U)^2
-(s_{\theta}^U)^2]s_{\alpha}}{\sqrt{2}}
H\bar{U}_{1}U_{2} 
\nonumber\\
&&
+\frac{iY_Uc_{\beta}}{\sqrt{2}}
A\bar{U}_{\text{2}}U_{\text{1}}
-\frac{iY_Uc_{\beta}}{\sqrt{2}}
A\bar{U}_{\text{1}}U_{\text{2}}. 
\end{eqnarray}
For the down-type VLFs, the Yukawa
Lagrangian is expanded as shown below:
\begin{eqnarray}
-\mathcal{L}^{\textrm{VLF}}_Y
(\phi_j D\bar{D} )
&=&
\sqrt{2}Y_D s_{\theta}^Dc_{\theta}^Ds_{\alpha}
h\bar{D}_{\text{1}}D_{\text{1}}
-\sqrt{2}Y_D s_{\theta}^Dc_{\theta}^Ds_{\alpha}
h\bar{D}_{\text{2}}D_{\text{2}}
+\frac{Y_D[(s_{\theta}^D)^2
-(c_{\theta}^D)^2]s_{\alpha}}{\sqrt{2}}
h\bar{D}_{2}D_{1}
\nonumber\\
&&
+\frac{Y_D[(s_{\theta}^D)^2
-(c_{\theta}^D)^2]s_{\alpha}}{\sqrt{2}}
h\bar{D}_{1}D_{2} 
-\sqrt{2}Y_D s_{\theta}^Dc_{\theta}^Dc_{\alpha}
H\bar{D}_{\text{1}}D_{\text{1}}
\nonumber\\
&&
+\sqrt{2}Y_D s_{\theta}^Dc_{\theta}^Dc_{\alpha}
H\bar{D}_{\text{2}}D_{\text{2}}
+\frac{Y_D[(c_{\theta}^D)^2
-(s_{\theta}^D)^2]c_{\alpha}}{\sqrt{2}}
H\bar{D}_{2}D_{1}
\nonumber\\
&&
+\frac{Y_D[(c_{\theta}^D)^2
-(s_{\theta}^D)^2]c_{\alpha}}{\sqrt{2}}
H\bar{D}_{1}D_{2}
+\frac{iY_Ds_{\beta}}{\sqrt{2}}
A\bar{D}_{\text{2}}D_{\text{1}}
-\frac{iY_Ds_{\beta}}{\sqrt{2}}
A\bar{D}_{\text{1}}D_{\text{2}}.
\end{eqnarray}
The interactions between the
charged Higgs and the VLFs are
formulated as:
\begin{eqnarray}
-\mathcal{L}^{\textrm{VLF}}_Y
(H^\pm UD )
&=&
(Y_Ds_{\beta}s_{\theta}^Dc_{\theta}^U
+Y_Uc_{\beta}s_{\theta}^Uc_{\theta}^D)
H^{\pm}\bar{U}_1D_1
+(Y_Uc_{\beta}s^U_{\theta}s^D_{\theta}
-Y_Ds_{\beta}c^U_{\theta}c^D_{\theta})
H^{\pm}\bar{U}_{\text{1}}D_{\text{2}}
\nonumber\\
&&
+(Y_Ds_{\beta}s^U_{\theta}s^D_{\theta}
-Y_Uc_{\beta}c^U_{\theta}c^D_{\theta})
H^{\pm}\bar{U}_{\text{2}}D_{\text{1}}
-
(Y_Uc_{\beta}s_{\theta}^Dc_{\theta}^U
+Y_Ds_{\beta}s_{\theta}^Uc_{\theta}^D)
H^{\pm}\bar{U}_{\text{2}}D_{\text{2}}
\nonumber\\
&&
+H.c.
\end{eqnarray}
By expanding the kinetic Lagrangian for the VLF model,
the interactions between the VLFs and gauge bosons
can be determined. Specifically, the resulting couplings
are given by:
\begin{eqnarray}
\mathcal{L}^{\textrm{VLF}}_{K}
&=&
i\bar{\mathcal{Q}}_{L}\gamma^{\mu}D_{\mu}\mathcal{Q}_{L}
+i\bar{\mathcal{Q}}_{R}\gamma^{\mu}D_{\mu}\mathcal{Q}_{R}
+i\bar{U}_{L,R}\gamma^{\mu}D_{\mu}{U}_{L,R}
+i\bar{D}_{L,R}\gamma^{\mu}D_{\mu}{D}_{L,R} 
\nonumber\\
&\supset&
\frac{g}{\sqrt{2}}
\Big[
c_{\theta}^{U}c_{\theta}^{D}\bar{U}_1\gamma^{\mu}D_1
+s_{\theta}^{U}s_{\theta}^{D}\bar{U}_2\gamma^{\mu}D_2
+c_{\theta}^{U}s_{\theta}^{D}\bar{U}_1\gamma^{\mu}D_2
+s_{\theta}^{U}c_{\theta}^{D}\bar{U}_2\gamma^{\mu}D_1
\Big]
W^{\pm}_{\mu}
+ H.c.
\nonumber\\
&&
+
\Big[
\frac{g}{c_W}(I_3^{D}-Q_{D}s^2_W)
(c_{\theta}^{D})^2
-\frac{e}{c_W}Q_{D}s_W(s_{\theta}^{D})^2
\Big]
Z_{\mu}\bar{D}_1
\gamma^{\mu}D_1
\nonumber\\
&&
+\Big[
\frac{g}{c_W}
(I_3^{D}-Q_{D}s^2_W)
(s_{\theta}^{D})^2
-\frac{e}{c_W}Q_{D}s_W
(c_{\theta}^{D})^2
\Big]
Z_{\mu}\bar{D}_2\gamma^{\mu}D_2
\nonumber\\
&&
+
\Big[
\frac{g}{c_W}(I_3^{U}
-Q_{U}s^2_W)(c_{\theta}^{U})^2
-\frac{e}{c_W}Q_{U}s_W(s_{\theta}^{U})^2
\Big]
Z_{\mu}\bar{U}_1\gamma^{\mu}U_1 
\nonumber\\
&&
+
\Big[
\frac{g}{c_W}(I_3^{U}
-Q_{U}s^2_W)(s_{\theta}^{U})^2
-\frac{e}{c_W}Q_{U}s_W(c_{\theta}^{U})^2
\Big]
Z_{\mu}\bar{U}_2\gamma^{\mu}U_2
\nonumber\\
&&
+\Big[
\frac{g}{c_W}(I_3^{D}-Q_{D}s^2_W)
+
\frac{e}{c_W}Q_{D}s_W
\Big]
s_{\theta}^{D}c_{\theta}^{D}
Z_{\mu}
(\bar{D}_1\gamma^{\mu}D_2
+\bar{D}_2\gamma^{\mu}D_1)
\nonumber\\
&&
+\Big[
\frac{g}{c_W}
(I_3^{U}-Q_{U}s^2_W)
+\frac{e}{c_W}Q_{U}s_W
\Big]
s_{\theta}^{U}c_{\theta}^{U}
Z_{\mu}
(\bar{U}_1\gamma^{\mu}U_2
+\bar{U}_2\gamma^{\mu}U_1)
\nonumber\\
&&
+eQ_{U}A_{\mu}\bar{U}_1\gamma^{\mu}U_1
+eQ_{U}A_{\mu}\bar{U}_2\gamma^{\mu}U_2
+eQ_{D}A_{\mu}\bar{D}_1\gamma^{\mu}D_1
+eQ_{D}A_{\mu}\bar{D}_2\gamma^{\mu}D_2.
\end{eqnarray}  
We highlight that the couplings obtained
for the THDM-II-VLF framework in this study
serve as our first results.
\section*{Appendix C: Triplet-Higgs Models}
The Triplet-Higgs Model is the second example considered in this work, with a detailed summary provided in this appendix. In this model, an additional real Higgs triplet, denoted as $\Delta$, is introduced into the scalar sector with a hypercharge of $Y_{\Delta} = 2$.
For a thorough discussion of this model, we recommend the reader consult
Refs.~\cite{Chun:2012jw, Chen:2013dh, Arhrib:2011uy, Arhrib:2011vc, Akeroyd:2012ms, Akeroyd:2011zza, Akeroyd:2011ir, Aoki:2011pz, Kanemura:2012rs, Chabab:2014ara, Han:2015hba, Chabab:2015nel, Ghosh:2017pxl, Ashanujjaman:2021txz, Zhou:2022mlz}, where various phenomenological aspects are also discussed in detail.
The Higgs potential, which is both renormalizable and gauge-invariant, takes the following form:
\begin{eqnarray}
\mathcal{V}^{\textrm{THM}}(\Phi, \Delta)
&=& -m_{\Phi}^2{\Phi^{\dagger}\Phi}
+\frac{\lambda}{4}(\Phi^{\dagger}\Phi)^2 
+ M_{\Delta}^2 \textrm{Tr}
(\Delta^{\dagger} \Delta)
+[\mu(\Phi^T{i}\sigma^2\Delta^{\dagger}\Phi)
+{\rm H.c.}]\nonumber\\
&&+
\lambda_1(\Phi^{\dagger}\Phi) 
\textrm{Tr}(\Delta^{\dagger}{\Delta})
+\lambda_2( \textrm{Tr}\Delta^{\dagger}{\Delta})^2
+\lambda_3 \textrm{Tr}(\Delta^{\dagger}{\Delta})^2+ 
\lambda_4{\Phi^\dagger\Delta\Delta^{\dagger}\Phi},
\end{eqnarray}
where $\sigma^2$ represents the Pauli matrix,
and all Higgs self-couplings $\lambda_i$
($i = \overline{i,4}$) are real.
For EWSB, the two Higgs multiplets
are parameterized in the following manner:
\begin{eqnarray}
\Delta &=\begin{bmatrix}
{\delta^+ \over \sqrt{2}} &  \delta^{++} \\
\frac{1}{\sqrt{2} } 
(v_{\Delta} +\eta_{\Delta} 
+i \chi_{\Delta})
& -{\delta^+ \over \sqrt{2}}
\end{bmatrix}
\quad 
{\rm and}
\quad 
\Phi=
\begin{bmatrix}
\phi^+ \\
\frac{1}{\sqrt{2} } 
(v_{\Phi} +\eta_{\Phi} 
+i \chi_{\Phi})
\end{bmatrix}.
\label{representa-htm}
\end{eqnarray}
Here, $v_{\Phi}$ and $v_{\Delta}$ represent the VEVs of the two neutral Higgs components. The electroweak scale is defined as $v= \sqrt{v_{\Phi}^2+2v_{\Delta}^2} = 246$ GeV to ensure consistency with the SM.
After EWSB, the physical Higgs content
of the THM includes two pairs of charged Higgs bosons\textemdash{}doubly charged ($H^{\pm\pm}$) and singly charged ($H^\pm$)\textemdash{}along with a neutral CP-odd Higgs ($A$) and two CP-even Higgs bosons ($H$ and $h$), where $h$ corresponds to the SM-like Higgs boson. The transformation between the mass and flavor bases is given by:
\begin{eqnarray}
\begin{pmatrix}
\phi^{\pm}\\
 \delta^{\pm}
\end{pmatrix}
&=&
\begin{pmatrix}
c_{\beta^{\pm}} & 
-s_{\beta^{\pm}} \\
s_{\beta^{\pm}}
& c_{\beta^{\pm}}
\end{pmatrix}
\begin{pmatrix}
G^{\pm}\\
H^{\pm}
\end{pmatrix},
\\\begin{pmatrix}
\eta_{\Phi}\\
 \eta_{\Delta}
\end{pmatrix}
&=&
\begin{pmatrix}
c_{\alpha} &  -s_{\alpha} \\
s_{\alpha}
& c_{\alpha}
\end{pmatrix}
\begin{pmatrix}
h\\
H
\end{pmatrix},
\end{eqnarray}
and 
\begin{eqnarray}
\begin{pmatrix}
\chi_{\Phi}\\
 \chi_{\Delta}
\end{pmatrix}
=
\begin{pmatrix}
c_{\beta^{0}} & 
-s_{\beta^{0}} \\
s_{\beta^{0}}
& c_{\beta^{0}}
\end{pmatrix}
\begin{pmatrix}
G^{0}\\
A
\end{pmatrix},
\end{eqnarray}
where  $t_{\beta^0} = \sqrt{2} t_{\beta^{\pm}}$ with
$t_{\beta^{\pm}} = \frac{\sqrt{2} v_{\Delta}}{v_{\Phi}}$,
and the mixing angle $\alpha$ between the two
neutral Higgs fields is taken into account.
Their masses depend on the Higgs
self-couplings and $\mu$, and are expressed as:
\begin{eqnarray}
&& M_{H^{\pm\pm}}^2  =  \frac{\sqrt{2}\mu{v_{\Phi}^2}-\lambda_4v_{\Phi}^2v_\Delta-2\lambda_3v_\Delta^3}{2v_\Delta},
\\
&& M_{H^{\pm}}^2 =  \frac{(v_{\Phi}^2+2v_\Delta^2)\,[2\sqrt{2}\mu-\lambda_4v_\Delta]}{4v_\Delta},\\
&& M_{A}^2 =  \frac{\mu(v_{\Phi}^2+4v_\Delta^2)}{\sqrt{2}v_\Delta},
\\
&& M_H^2 = \frac{1}{2}\Big\{\lambda v_{\Phi}^2 s_\alpha^2 + c_\alpha^2 \Big[\sqrt{2}\mu\frac{v_{\Phi}^2}{v_\Delta}\big(1+4\frac{v_\Delta}{v_{\Phi}} t_{\alpha}\big) +4v_\Delta^2\big(\lambda_{23} - \lambda_{14}\frac{v_{\Phi}}{v_\Delta} t_{\alpha}\big) \Big]\Big\},
\\
&& M_h^2 = \frac{1}{2}
\Big\{
\lambda v_{\Phi}^2 c_\alpha^2 + s_\alpha^2 \Big[ \sqrt{2}\mu\frac{v_{\Phi}^2}{v_\Delta}\big(1-4\frac{v_\Delta}{v_{\Phi} t_{\alpha}}\big) +4v_\Delta^2\big( \lambda_{14} \frac{v_{\Phi}}{v_\Delta t_{\alpha}} +
\lambda_{23}
\big)\Big]
\Big\}.
\end{eqnarray}
In terms of the mass eigenstates,
the Yukawa Lagrangian is formulated as follows:
\begin{eqnarray}
\label{YukawaTHM}
{\mathcal{L}}^{\textrm{THM}}_{\rm Yukawa} =
{\mathcal{L}}^{\rm SM}_{\rm Yukawa}
- L^T y_{\nu}C (i \sigma^2 \Delta) L 
+ H.c,
\end{eqnarray}
where $L=(L_e,L_{\mu},L_{\tau})$
corresponds to the three left-handed
lepton doublets, $y_{\nu}$
denotes the $3\times3$ Yukawa
coupling matrix responsible for
neutrino mass generation, and $C$
is the charge conjugation operator.

The couplings between the scalar particles,
derived from the Higgs potential,
are expressed as follows:
\begin{eqnarray}
-
\mathcal{V}^{\textrm{THM}}
(\Phi, \Delta)
&\supset&
\bigg
\{
-\dfrac{\lambda
\; v_{\Phi}
\; c_{\alpha}
\; s_{\beta^\pm}^2
}
{2}
-
\lambda_1
(
s_{\alpha}
\;
s_{\beta^\pm}^2
\;
v_{\Delta}
+
c_{\alpha}
\;
c_{\beta^\pm}^2
\;
v_{\Phi}
)
-
2
\lambda_{23}
\;
s_{\alpha}
\;
c_{\beta^\pm}^2
\;
v_{\Delta}
\nonumber\\
&&
\hspace{0.5cm}
-2
c_{\alpha}
c_{\beta^\pm}
s_{\beta^\pm}
\;
\mu
-
\frac{\lambda_4}{2}
\Big[
v_{\Phi}
c_{\alpha}
c_{\beta^\pm}^2
-
\frac{
s_{2\beta^\pm}
}
{
\sqrt{2}
}
(
s_{\alpha}
v_{\Phi}
+
c_{\alpha}
v_{\Delta}
)
\Big]
\bigg
\}
hH^{\pm}H^{\mp} 
\nonumber\\
&&
+
\bigg
\{
\frac{\lambda}{2}
s_{\beta^\pm}^2
v_{\Phi}
s_{\alpha}
+
\lambda_1
(c_{\beta^\pm}^2
v_{\Phi}
s_{\alpha}-
s_{\beta^\pm}^2
v_{\Delta}
c_{\alpha})
-
2\lambda_{23}
\;
c_{\alpha}
\;
c_{\beta^\pm}^2
\;
v_{\Delta}
\nonumber\\
&&
\hspace{0.5cm}
+2 s_{\alpha}
c_{\beta^\pm}
s_{\beta^\pm}
\mu
+
\frac{\lambda_4}{2}   
[v_{\Phi}
s_{\alpha}
c_{\beta^\pm}^2
+\sqrt{2}
s_{\beta^\pm}
c_{\beta^\pm}
(v_{\Phi}
c_{\alpha}
-
v_{\Delta}
s_{\alpha})]
\bigg
\}
HH^{\pm}H^{\mp}
\nonumber\\
&&
-
\Big
(\lambda_1
v_{\Phi}
c_{\alpha}
+
2\lambda_2
v_{\Delta}
s_{\alpha}
\Big)
hH^{\pm\pm}H^{\mp\mp}
+
\Big(
\lambda_1
v_{\Phi}
s_{\alpha}
-
2\lambda_2
v_{\Delta}
c_{\alpha}
\Big)
HH^{\pm\pm}H^{\mp\mp}
\nonumber\\
&&+
\cdots
.
\end{eqnarray}
In the limit $\frac{v_{\Delta} }{v_{\Phi} } \rightarrow 0$, the parameter $s_{\beta^{\pm}}$ approaches zero, yielding the following result:
\begin{eqnarray}
g_{hH^{\pm}H^{\mp}}^{\textrm{THM} }
&=&
-\lambda_1 c_{\alpha} v
-
\frac{\lambda_4}{2} c_{\alpha} v, \\
g_{HH^{\pm}H^{\mp}}^{\textrm{THM} }
&=&
\lambda_1 s_{\alpha} v
+
\frac{\lambda_4}{2}
s_{\alpha} v, \\
g_{hH^{\pm\pm}H^{\mp\mp}}^{\textrm{THM} }
&=&
-\lambda_1\;c_{\alpha}\;v, \\
g_{HH^{\pm\pm}H^{\mp\mp}}^{\textrm{THM} }
&=&
\lambda_1\; s_{\alpha}\;v.
\end{eqnarray}
Where $\lambda_4 
=4\frac{M_A^2-M_{H^\pm}^2 }{v^2}$
is taken into account 
in the above couplings.

The total decay width
of CP-even Higgs
$H$ is calculated at LO
as in~\cite{Kanemura:2022ldq}
\begin{eqnarray}
\label{GMAH}
\Gamma_H &\approx& \sum
\limits_{f}
\frac{N_C^f\; M_H}{8\pi}
\left(\kappa_H^{f}
\frac{m_f}{v} \right)^2
\left(
1 - \frac{4m_f^2}{M_H^2}
\right)^{3/2}
\\
&&
+\sum\limits_{V=Z, W}
\frac{M_H^3}{64\pi\; c_V \; M_V^4}
\left(\kappa_H^V
\frac{2M_V^2}{v}
\right)^2
\left(
 1-\frac{4M_V^2}{M_H^2} +
\frac{12M_V^4}{M_H^4}
\right)
\sqrt{1-\frac{4M_V^2}{M_H^2} }
.
\nonumber
\end{eqnarray}
Where $c_V=1(2)$ for $W (Z)$
respectively. The coefficient couplings
$\kappa_H^{f}
\sim c_{\beta -\alpha}$,
$\kappa_H^{V} =c_{\beta-\alpha}$ for THDM
and
$\kappa_H^{f}= c_{\alpha}$,
$\kappa_H^{V}
= c_{\alpha}\; c_{\beta^{\pm}}
+\sqrt{2} s_{\alpha}\; s_{\beta^{\pm}}$
for THM, respectively.




\begin{thebibliography}{100}
\bibitem{Dermisek:2009fd}
R.~Dermisek and J.~F.~Gunion,
Phys. Rev. D \textbf{81} (2010), 055001
doi:10.1103/PhysRevD.81.055001
[arXiv:0911.2460 [hep-ph]].
\bibitem{ATLAS:2015kpj}
G.~Aad \textit{et al.} [ATLAS],
Phys. Lett. B \textbf{744} (2015), 163-183
doi:10.1016/j.physletb.2015.03.054
[arXiv:1502.04478 [hep-ex]].
\bibitem{CMS:2020ffa}
A.~M.~Sirunyan \textit{et al.} [CMS],
JHEP \textbf{08} (2020), 139
doi:10.1007/JHEP08(2020)139
[arXiv:2005.08694 [hep-ex]].
\bibitem{ATLAS:2024rzd}
G.~Aad \textit{et al.} [ATLAS],
[arXiv:2409.20381 [hep-ex]].
\bibitem{CMS:2012fgd}
S.~Chatrchyan \textit{et al.} [CMS],
Phys. Rev. Lett. \textbf{109} (2012), 121801
doi:10.1103/PhysRevLett.109.121801
[arXiv:1206.6326 [hep-ex]].

\bibitem{ATLAS:2023zkt}
G.~Aad \textit{et al.} [ATLAS],
JHEP \textbf{02} (2024), 197
doi:10.1007/JHEP02(2024)197
[arXiv:2311.04033 [hep-ex]].
\bibitem{Aiko:2022gmz}
M.~Aiko, S.~Kanemura and K.~Sakurai,
Nucl. Phys. B \textbf{986} (2023), 116047
doi:10.1016/j.nuclphysb.2022.116047
[arXiv:2207.01032 [hep-ph]].
\bibitem{Kanemura:2017gbi}
S.~Kanemura, M.~Kikuchi, K.~Sakurai and K.~Yagyu,
Comput. Phys. Commun. \textbf{233} (2018), 134-144
doi:10.1016/j.cpc.2018.06.012
[arXiv:1710.04603 [hep-ph]].

\bibitem{Kanemura:2019slf}
S.~Kanemura, M.~Kikuchi, K.~Mawatari, K.~Sakurai and K.~Yagyu,
Comput. Phys. Commun. \textbf{257} (2020), 107512
doi:10.1016/j.cpc.2020.107512
[arXiv:1910.12769 [hep-ph]].

\bibitem{Aiko:2023xui}
M.~Aiko, S.~Kanemura, M.~Kikuchi, K.~Sakurai and K.~Yagyu,
Comput. Phys. Commun. \textbf{301} (2024), 109231
doi:10.1016/j.cpc.2024.109231
[arXiv:2311.15892 [hep-ph]].

\bibitem{Akeroyd:2024tbp}
A.~G.~Akeroyd, S.~Alanazi and S.~Moretti,
[arXiv:2408.08314 [hep-ph]].
\bibitem{Akeroyd:2023kek}
A.~G.~Akeroyd, S.~Alanazi and S.~Moretti,
J. Phys. G \textbf{50} (2023) no.9, 095001
doi:10.1088/1361-6471/ace3e1
[arXiv:2301.00728 [hep-ph]].
\bibitem{Weber:2003tw}
C.~Weber, H.~Eberl and W.~Majerotto,
Phys. Rev. D \textbf{68} (2003), 093011
doi:10.1103/PhysRevD.68.093011
[arXiv:hep-ph/0308146 [hep-ph]].
\bibitem{Weber:2003eg}
C.~Weber, H.~Eberl and W.~Majerotto,
Phys. Lett. B \textbf{572} (2003), 56-67
doi:10.1016/j.physletb.2003.07.083
[arXiv:hep-ph/0305250 [hep-ph]].
\bibitem{Chishtie:1999dd}
F.~A.~Chishtie, V.~Elias and T.~G.~Steele,
J. Phys. G \textbf{26} (2000), 93-98
doi:10.1088/0954-3899/26/1/309
[arXiv:hep-ph/9909374 [hep-ph]].

\bibitem{Arhrib:2018pdi}
A.~Arhrib, R.~Benbrik, J.~El Falaki, M.~Sampaio and R.~Santos,
Phys. Rev. D \textbf{99} (2019) no.3, 035043
doi:10.1103/PhysRevD.99.035043
[arXiv:1809.04805 [hep-ph]].

\bibitem{Arco:2025ydq}
F.~Arco, T.~Biek\"otter, P.~Stylianou and G.~Weiglein,
[arXiv:2502.03443 [hep-ph]].

\bibitem{Banik:2024ugs}
S.~Banik, G.~Coloretti, A.~Crivellin and H.~E.~Haber,
[arXiv:2412.00523 [hep-ph]].

\bibitem{Phan:2024zus}
K.~H.~Phan, D.~T.~Tran and T.~H.~Nguyen,
PTEP \textbf{2024} (2024) no.8, 083B02
doi:10.1093/ptep/ptae103
[arXiv:2404.02417 [hep-ph]].
\bibitem{Tran:2025dea}
D.~T.~Tran, L.~T.~Hue, T.~H.~Nguyen, V.~Q.~Phong and K.~H.~Phan,
[arXiv:2501.15239 [hep-ph]].

\bibitem{Sanchez-Velez:2018xdj}
R.~S\'anchez-V\'elez and G.~Tavares-Velasco,
Phys. Rev. D \textbf{97} (2018) no.9, 095038
doi:10.1103/PhysRevD.97.095038
[arXiv:1802.01222 [hep-ph]].

\bibitem{Sanchez-Velez:2019nsh}
R.~S\'anchez-V\'elez and G.~Tavares-Velasco,
Phys. Rev. D \textbf{99} (2019) no.5, 055024
doi:10.1103/PhysRevD.99.055024
[arXiv:1901.05399 [hep-ph]].


\bibitem{Yin:2002sq}
J.~Yin, W.~G.~Ma, R.~Y.~Zhang and H.~S.~Hou,
Phys. Rev. D \textbf{66} (2002), 095008
doi:10.1103/PhysRevD.66.095008.

\bibitem{Akeroyd:1999gu}
A.~G.~Akeroyd, A.~Arhrib and M.~Capdequi Peyranere,
Mod. Phys. Lett. A \textbf{14} (1999), 2093-2108
[erratum: Mod. Phys. Lett. A \textbf{17} (2002), 373]
doi:10.1142/S0217732399002157
[arXiv:hep-ph/9907542 [hep-ph]].


\bibitem{Akeroyd:2001aka}
A.~G.~Akeroyd, A.~Arhrib and M.~Capdequi Peyranere,
Phys. Rev. D \textbf{64} (2001), 075007
[erratum: Phys. Rev. D \textbf{65} (2002), 099903]
doi:10.1103/PhysRevD.65.099903
[arXiv:hep-ph/0104243 [hep-ph]].


\bibitem{Sasaki:2017fvk}
K.~Sasaki and T.~Uematsu,
Phys. Lett. B \textbf{781} (2018), 290-294
doi:10.1016/j.physletb.2018.04.005
[arXiv:1712.00197 [hep-ph]].

\bibitem{Farris:2003pn}
T.~Farris, J.~F.~Gunion, H.~E.~Logan and S.~f.~Su,
Phys. Rev. D \textbf{68} (2003), 075006
doi:10.1103/PhysRevD.68.075006
[arXiv:hep-ph/0302266 [hep-ph]].

\bibitem{Arhrib:2002ti}
A.~Arhrib,
Phys. Rev. D \textbf{67} (2003), 015003
doi:10.1103/PhysRevD.67.015003
[arXiv:hep-ph/0207330 [hep-ph]].


\bibitem{Denner:1991kt}
A.~Denner,
Fortsch. Phys. \textbf{41} (1993), 307-420
doi:10.1002/prop.2190410402
[arXiv:0709.1075 [hep-ph]].

\bibitem{Hahn:1998yk}
T.~Hahn and M.~Perez-Victoria,
Comput. Phys. Commun. \textbf{118} (1999), 153-165
doi:10.1016/S0010-4655(98)00173-8
[arXiv:hep-ph/9807565 [hep-ph]].


\bibitem{Denner:2016kdg}
A.~Denner, S.~Dittmaier and L.~Hofer,
Comput. Phys. Commun. \textbf{212} (2017), 220-238
doi:10.1016/j.cpc.2016.10.013
[arXiv:1604.06792 [hep-ph]].

\bibitem{Hahn:2000kx}
T.~Hahn,
Comput. Phys. Commun. \textbf{140} (2001), 418-431
doi:10.1016/S0010-4655(01)00290-9
[arXiv:hep-ph/0012260 [hep-ph]].

\bibitem{Hahn:2016ebn}
T.~Hahn, S.~Pa\ss{}ehr and C.~Schappacher,
PoS \textbf{LL2016} (2016), 068
doi:10.1088/1742-6596/762/1/012065
[arXiv:1604.04611 [hep-ph]].
\bibitem{Branco:2011iw}
G.~C.~Branco, P.~M.~Ferreira, L.~Lavoura, M.~N.~Rebelo, M.~Sher and J.~P.~Silva,
Phys. Rept. \textbf{516} (2012), 1-102
doi:10.1016/j.physrep.2012.02.002
[arXiv:1106.0034 [hep-ph]].

\bibitem{Aoki:2009ha}
M.~Aoki, S.~Kanemura, K.~Tsumura and K.~Yagyu,
Phys. Rev. D \textbf{80} (2009), 015017
doi:10.1103/PhysRevD.80.015017
[arXiv:0902.4665 [hep-ph]].

\bibitem{Song:2019aav}
J.~Song and Y.~W.~Yoon,
Phys. Rev. D \textbf{100} (2019) no.5, 055006
doi:10.1103/PhysRevD.100.055006
[arXiv:1904.06521 [hep-ph]].

\bibitem{Chun:2012jw}
E.~J.~Chun, H.~M.~Lee and P.~Sharma,
JHEP \textbf{11} (2012), 106
doi:10.1007/JHEP11(2012)106
[arXiv:1209.1303 [hep-ph]].
\bibitem{Chen:2013dh}
C.~S.~Chen, C.~Q.~Geng, D.~Huang and L.~H.~Tsai,
Phys. Lett. B \textbf{723} (2013), 156-160
doi:10.1016/j.physletb.2013.05.007
[arXiv:1302.0502 [hep-ph]].
\bibitem{Arhrib:2011uy}
A.~Arhrib, R.~Benbrik, M.~Chabab, G.~Moultaka, M.~C.~Peyranere, L.~Rahili and J.~Ramadan,
Phys. Rev. D \textbf{84} (2011), 095005
doi:10.1103/PhysRevD.84.095005
[arXiv:1105.1925 [hep-ph]].
\bibitem{Arhrib:2011vc}
A.~Arhrib, R.~Benbrik, M.~Chabab, G.~Moultaka and L.~Rahili,
JHEP \textbf{04} (2012), 136
doi:10.1007/JHEP04(2012)136
[arXiv:1112.5453 [hep-ph]].
\bibitem{Akeroyd:2012ms}
A.~G.~Akeroyd and S.~Moretti,
Phys. Rev. D \textbf{86} (2012), 035015
doi:10.1103/PhysRevD.86.035015
[arXiv:1206.0535 [hep-ph]].
\bibitem{Akeroyd:2011zza}
A.~G.~Akeroyd and H.~Sugiyama,
Phys. Rev. D \textbf{84} (2011), 035010
doi:10.1103/PhysRevD.84.035010
[arXiv:1105.2209 [hep-ph]].
\bibitem{Akeroyd:2011ir}
A.~G.~Akeroyd and S.~Moretti,
Phys. Rev. D \textbf{84} (2011), 035028
doi:10.1103/PhysRevD.84.035028
[arXiv:1106.3427 [hep-ph]].
\bibitem{Aoki:2011pz}
M.~Aoki, S.~Kanemura and K.~Yagyu,
Phys. Rev. D \textbf{85} (2012), 055007
doi:10.1103/PhysRevD.85.055007
[arXiv:1110.4625 [hep-ph]].
\bibitem{Kanemura:2012rs}
S.~Kanemura and K.~Yagyu,
Phys. Rev. D \textbf{85} (2012), 115009
doi:10.1103/PhysRevD.85.115009
[arXiv:1201.6287 [hep-ph]].
\bibitem{Chabab:2014ara}
M.~Chabab, M.~C.~Peyranere and L.~Rahili,
Phys. Rev. D \textbf{90} (2014) no.3, 035026
doi:10.1103/PhysRevD.90.035026
[arXiv:1407.1797 [hep-ph]].
\bibitem{Han:2015hba}
Z.~L.~Han, R.~Ding and Y.~Liao,
Phys. Rev. D \textbf{91} (2015), 093006
doi:10.1103/PhysRevD.91.093006
[arXiv:1502.05242 [hep-ph]].
\bibitem{Chabab:2015nel}
M.~Chabab, M.~C.~Peyran\`ere and L.~Rahili,
Phys. Rev. D \textbf{93} (2016) no.11, 115021
doi:10.1103/PhysRevD.93.115021
[arXiv:1512.07280 [hep-ph]].
\bibitem{Haba:2016zbu}
N.~Haba, H.~Ishida, N.~Okada and Y.~Yamaguchi,
Eur. Phys. J. C \textbf{76} (2016) no.6, 333
doi:10.1140/epjc/s10052-016-4180-z
[arXiv:1601.05217 [hep-ph]].
\bibitem{Ghosh:2017pxl}
D.~K.~Ghosh, N.~Ghosh, I.~Saha and A.~Shaw,
Phys. Rev. D \textbf{97} (2018) no.11, 115022
doi:10.1103/PhysRevD.97.115022
[arXiv:1711.06062 [hep-ph]].
\bibitem{Ashanujjaman:2021txz}
S.~Ashanujjaman and K.~Ghosh,
JHEP \textbf{03} (2022), 195
doi:10.1007/JHEP03(2022)195
[arXiv:2108.10952 [hep-ph]].
\bibitem{Zhou:2022mlz}
R.~Zhou, L.~Bian and Y.~Du,
JHEP \textbf{08} (2022), 205
doi:10.1007/JHEP08(2022)205
[arXiv:2203.01561 [hep-ph]].
\bibitem{Aoki:2012jj}
M.~Aoki, S.~Kanemura, M.~Kikuchi and K.~Yagyu,
Phys. Rev. D \textbf{87} (2013) no.1, 015012
doi:10.1103/PhysRevD.87.015012
[arXiv:1211.6029 [hep-ph]].
\bibitem{Kanemura:2022ldq}
S.~Kanemura, M.~Kikuchi and K.~Yagyu,
Nucl. Phys. B \textbf{983} (2022), 115906
doi:10.1016/j.nuclphysb.2022.115906
[arXiv:2203.08337 [hep-ph]].
\bibitem{Hahn:2004fe}
T.~Hahn,
Comput. Phys. Commun. \textbf{168} (2005), 78-95
doi:10.1016/j.cpc.2005.01.010
[arXiv:hep-ph/0404043 [hep-ph]].
\bibitem{Hue:2023tdz}
L.~T.~Hue, D.~T.~Tran, T.~H.~Nguyen and K.~H.~Phan,
PTEP \textbf{2023} (2023) no.8, 083B06
doi:10.1093/ptep/ptad106
[arXiv:2305.04002 [hep-ph]].
\bibitem{Phan:2024vfy}
K.~H.~Phan, D.~T.~Tran and T.~H.~Nguyen,
[arXiv:2409.00662 [hep-ph]].
\bibitem{Phan:2024jbx}
K.~H.~Phan, D.~T.~Tran and T.~H.~Nguyen,
PTEP \textbf{2025} (2025) no.2, 023B05
doi:10.1093/ptep/ptaf012
[arXiv:2406.15749 [hep-ph]].
\end{thebibliography}
\end{document}